\documentclass[a4paper,12pt]{article}
\pdfoutput=1

\usepackage{amssymb,amsmath,bm}
\usepackage{a4wide}
\usepackage{color}
\usepackage{slashed}
\usepackage{graphicx}
\usepackage{amsfonts}
\usepackage{lscape}
\usepackage{hyperref}
\usepackage{amsthm}
\usepackage{tikz}
\usepackage{subfig}
\usepackage{multirow}
\usepackage[compat=1.1.0]{tikz-feynman}

\hypersetup{
   bookmarks=true,         
   unicode=true,          
   pdftoolbar=true,        
   pdfmenubar=true,        
   pdffitwindow=false,     
   pdfstartview={FitH},    
   pdftitle={My title},    
   pdfauthor={Author},     
   pdfsubject={Subject},   
   pdfcreator={Creator},   
   pdfproducer={Producer}, 
   pdfkeywords={keyword1} {key2} {key3}, 
   pdfnewwindow=true,      
   colorlinks=true,       
   linkcolor=blue,          
   citecolor=magenta,        
   filecolor=magenta,      
   urlcolor=cyan           
}

\usepackage{booktabs}
\usepackage{enumerate}
\usepackage{array}
\usepackage{rotating}
\usepackage[numbers, sort&compress]{natbib}
\usepackage{float}
\usepackage[utf8]{inputenc}
\usepackage[T1]{fontenc}
\usepackage[utf8]{inputenc}
\newcommand{\be}{\begin{equation}}
\newcommand{\ee}{\end{equation}}

\newcommand{\beq}{\begin{equation}} 
\newcommand{\eeq}{\end{equation}} 
\newcommand{\ba}{\begin{array}}  
\newcommand{\ea}{\end{array}} 
\newcommand{\bea}{\begin{eqnarray}}  
\newcommand{\eea}{\end{eqnarray} }  
\newcommand{\bal}{\begin{align}}
\newcommand{\eal}{\end{align}}   
\newcommand{\bi}{\begin{itemize}}  
\newcommand{\ei}{\end{itemize}}  
\newcommand{\ben}{\begin{enumerate}}  
\newcommand{\een}{\end{enumerate}}  
\newcommand{\bc}{\begin{center}}
\newcommand{\ec}{\end{center}} 
\newcommand{\bt}{\begin{table}}
\newcommand{\et}{\end{table}}  
\newcommand{\btb}{\begin{tabular}}
\newcommand{\etb}{\end{tabular}}

\newcommand{\MET}{E\llap{/\kern1.5pt}_T}

\newcommand{\slepton}{\phi}
\newcommand{\sfermion}{\phi}
\newcommand{\smuon}{\widetilde{\mu}}

\definecolor{Grn}{rgb}{0.,0.6,0.}

\begin{document}

\begin{titlepage}
\vspace*{-1.0truecm}
\begin{flushright}
ULB-TH/19-01\\
 \vspace*{2mm}
 \end{flushright}
\vspace{0.8truecm}

\begin{center}
\boldmath

\textbf{
{\LARGE  A feeble window on leptophilic dark matter
}}
\unboldmath
\end{center}

\vspace{0.4truecm}

\begin{center}
{\large
{\bf  Sam Junius$\,^{a,b,c}$, Laura Lopez-Honorez$\,^{a,b}$, 
\\[1.5mm]  and Alberto Mariotti$\,^{b,c}$}
}
\vspace{0.6truecm}

{\footnotesize
$^a${\sl Service de Physique Th\'eorique, Universit\'e Libre de Bruxelles, C.P. 225,  B-1050 Brussels, Belgium }
  $^b${\sl Theoretische Natuurkunde \& The International Solvay Institutes, Vrije Universiteit Brussel, Pleinlaan 2, B-1050 Brussels, Belgium \\}
  $^c${\sl  Inter-University Institute for High Energies, Vrije Universiteit Brussel, Pleinlaan 2, B-1050 Brussels, Belgium \\}

}
\end{center}

\vspace{0.4truecm}
\begin{abstract}
  \noindent
  In this paper we study a leptophilic dark matter scenario involving
  feeble dark matter coupling to the Standard Model (SM) and
  compressed dark matter-mediator mass spectrum.  We consider a
  simplified model where the SM is extended with one Majorana fermion,
  the dark matter, and one charged scalar, the mediator, coupling to
  the SM leptons through a Yukawa interaction. We first discuss the
  dependence of the dark matter relic abundance on the Yukawa coupling
  going continuously from freeze-in to freeze-out with an intermediate
  stage of conversion driven freeze-out.  Focusing on the latter, we
  then exploit the macroscopic decay length of the charged scalar to
  study the resulting long-lived-particle signatures at collider and
  to explore the experimental reach on the viable portion of the
  parameter space.

 \end{abstract}

\end{titlepage}
\tableofcontents

\section{Introduction}
\setcounter{equation}{0}

As of today, the most conventional paradigm for dark matter (DM) has
been the so-called Weakly Interacting Massive Particle (WIMP). WIMPs
have been the main target of experimental searches, including collider
experiments, direct and indirect DM detection.  In the WIMP scenario,
the DM is produced in the early Universe through the freeze-out
mechanism, leading typically to the correct DM relic abundance for
electroweak size couplings and masses. This is however not the only
possibility to obtain the right DM abundance.  By varying the DM mass,
its coupling strength to the Standard Model (SM) and/or within the
dark sector, one can generate DM through different mechanisms during
the cosmological evolution of the Universe.  Specifically, one can
continuously go from freeze-out to freeze-in passing through several
intermediate DM production regimes, see
e.g.~\cite{Chu:2011be,Bernal:2017kxu, Garny:2017rxs,Hall:2009bx}. In
some parts of this parameter space, the DM happens to be very feebly
coupled to the SM, i.e. with couplings much more suppressed than for
the WIMP case.

We focus on this feeble interaction window for scenarios involving a
small mass splitting between the dark matter and the mediator
connecting DM to the SM.  We will study in details the mechanism of
dark matter production in the early universe, from freeze-in to
freeze-out. In particular, we will mainly focus on the intermediate
stage of DM coannihilation freeze-out happening out of chemical
equilibrium (CE) with the SM plasma, also called conversion driven
freeze-out. Such a scenario has already been pointed out
in~\cite{Garny:2017rxs} and mainly studied for dark matter coupling to
quarks~\cite{Garny:2017rxs,Garny:2018icg,Garny:2019kua}. Here instead
we focus on the case of a leptophilic dark matter model. Conversion
processes between the mediator and the dark matter will play a central
role in defining the evolution of the DM abundance and they will have
to be taken into account in the study of the DM/mediator Boltzmann
equations. In passing, we will also emphasize the fact that, within
the freeze-in framework, mediator scatterings (as opposed to decay)
can play a leading role in determining the DM relic density. Finally,
we will identify the viable parameter space for conversion driven
freeze-out so as to further study the experimental constraints on this
class of models.

 The feeble coupling of the DM to the mediator allows for a
 macroscopic decay length of the mediator that can be observed at
 colliders through e.g. charged and/or disappearing tracks. These are
 typical features of DM scenarios in which the DM abundance results
 from the freeze-in~\cite{Hall:2009bx} and conversion driven
 freeze-out~\cite{Garny:2017rxs}. These production mechanisms can lead
 to distinctive and challenging signatures at colliders, including
 long lived charged particles and very soft signatures. In the
 freeze-in case, the DM coupling is so suppressed that the mediator
 mainly decays outside the detector giving rise to charged tracks. For
 conversion driven freeze-out, the slightly larger couplings involved
 can also give rise to disappearing tracks.  The LHC community has
 already provided a strong effort in the study of final state
 signatures which arise from DM models, focusing mainly on the WIMP
 scenario and on prompt missing energy signatures, see
 e.g. \cite{Abercrombie:2015wmb}.  Recently, more attention has been
 devoted to long lived particle signatures arising in DM models, that
 we will study here, see
 e.g.~\cite{Hall:2009bx,Co:2015pka,Hessler:2016kwm,DEramo:2017ecx,Davoli:2017swj,Calibbi:2018fqf,Garny:2018icg,Heisig:2018kfq,Garny:2017rxs,Brooijmans:2018xbu,Belanger:2018sti,Filimonova:2018qdc,Heisig:2018teh,Chang:2009sv,Buchmueller:2017uqu,Ghosh:2017vhe,Davoli:2018mau,Alimena:2019zri}.
 Notice that due to the feeble coupling involved, direct and indirect
 detection dark matter searches are challenging, see however
 e.g.~\cite{Chu:2011be,Chu:2016pew,Bernal:2015ova,
   Hochberg:2017wce,Knapen:2017ekk,Heikinheimo:2018duk,Bernal:2018ins,
   Hambye:2018dpi}.  Unconventional signatures at the LHC can hence
 provide the main experimental probes for the class of model studied
 here.

The rest of the paper is organized as follows. In
Sec.~\ref{sec:simpl-lept-dark}, we introduce the leptophilic DM model
on which we focus all along this work. In Sec.~\ref{sec:conv}, we
detail the computation of the dark matter relic abundance in different
regimes underlying the specificities of the compressed spectrum
scenarios.  In Sec.~\ref{sec:viab} we turn to the unexplored window on
leptophilic DM provided by the conversion driven mechanism,
determining the viable parameter space and the corresponding collider
constraints. We summarize and conclude in Sec.~\ref{sec:concl}, while
we present some technical details in the Appendices.

\section{The simplified leptophilic dark matter model}
\label{sec:simpl-lept-dark}

In this paper, we work in a minimal extension of the Standard Model (SM) involving a
Majorana fermion $\chi$ dark matter coupled to SM leptons
through the exchange of  charged scalar mediator $\phi$. The
Lagrangian encapsulating the BSM physics reads
\begin{equation}
  {\cal L}\supset \frac{1}{2}  \bar{\chi} \gamma^{\mu} \partial_{\mu} \chi-\frac{m_{\chi}}{2} \bar{\chi} \chi+ (D_{\mu}\sfermion)^\dagger \ D^{\mu}\sfermion -m_\phi^2 |\phi|^2 \ - \
  \lambda_{\chi} \sfermion  \bar \chi l_R \ + \ h.c.
\label{eq:lagr}
\end{equation}
where $m_{\chi} $ is the dark matter  mass, $m_\phi$ is the
mediator mass and $\lambda_\chi$ denotes the Yukawa coupling between
the dark matter, the right handed SM lepton $l_R$ and the mediator. We
have assumed that a $Z_2$ symmetry prevents the dark matter to decay
directly to SM particles.  Both $\chi$ and $\phi$ are odd under the
$Z_2$ symmetry while the SM fields are even and we assume
$m_\phi>m_\chi$.

\begin{figure}[t!]
	\centering
	\subfloat[]{\label{fig:feyn_decay}
			\begin{tikzpicture}
			\begin{feynman}[large]
			\vertex (a) ;
			\vertex[left=of a] (i1) {\(\sfermion\)};
			\vertex[above right=of a] (f1){\(\chi\)};
			\vertex[below right=of a] (f2){\(l\)};
			\diagram* {	
				(i1) -- [scalar] (a) -- (f1),
				(a) -- (f2),
			};
			\end{feynman}
			\end{tikzpicture}}
	\qquad
	\subfloat[]{\label{fig:feyn_scattering}
			\begin{tikzpicture}
			\begin{feynman}[large]
			\node[blob] (a);
			\vertex[above left=of a] (i1) {$\chi$};
			\vertex[below left=of a] (i2) {\(SM\)};
			\vertex[above right=of a] (f1){\(\sfermion\)};
			\vertex[below right=of a] (f2){\(SM\)};
			\diagram* {	
				(i1) -- (a) -- [scalar] (f1),
				(i2) -- (a) -- (f2),
			};
			\end{feynman}
			\end{tikzpicture}}
			\caption{The Feynman diagrams for the most important conversion processes.}
			\label{fig:feyn_conv}
\end{figure}
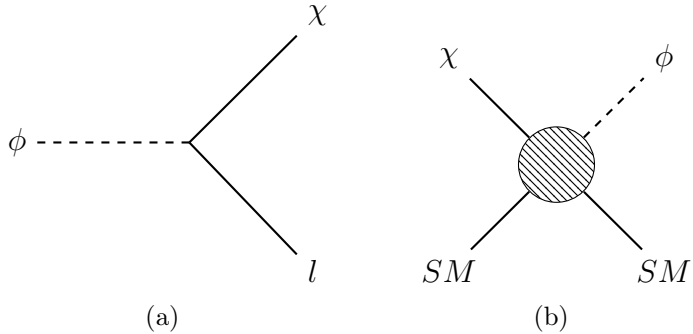

The standard WIMP phenomenology of (\ref{eq:lagr}) has already been
studied in length
in~\cite{Garny:2011ii,Garny:2015wea,Garny:2013ama,Bringmann:2012vr,Kopp:2014tsa,Baker:2018uox,Khoze:2017ixx}
while the corresponding scalar DM case has been studied
in~\cite{Giacchino:2013bta,Giacchino:2014moa,Giacchino:2015hvk,
  Colucci:2018vxz,Toma:2013bka,Ibarra:2014qma}.\footnote{See
  also~\cite{Belanger:2018sti} for the freeze-in case.}  Here in
contrast, we extend existing analysis to couplings $\lambda_\chi$
bringing the dark matter out of chemical equilibrium (CE) focusing on
small mass splittings between the dark matter and the mediator. In
this set-up, the conversion $\chi\leftrightarrow \phi$ processes
depicted in Fig.~\ref{fig:feyn_conv} will play the leading roles. We
will study how the phenomenology of the model (\ref{eq:lagr})
non-trivially evolves between the two extremes of freeze-in and
standard freeze-out.  For an interaction strength just below CE limit,
i.e. couplings $ \lambda_{\chi} \lesssim 10^{-6}$ (see
Sec.~\ref{sec:illustr-benchm-model}), the Majorana fermion can account
for all the dark matter through conversion driven
freeze-out~\cite{Garny:2018icg,Garny:2017rxs}. In the latter case,
suppressed conversion will allow for a larger dark matter abundance
than the one predicted when assuming CE. For even lower interaction
strength, i.e. couplings $ \lambda_{\chi} \lesssim 10^{-8}$, we enter
the freeze-in regime within which the dark matter abundance results
from the processes converting the mediator to the dark matter
$\phi\rightarrow \chi$, see e.g.~\cite{Hall:2009bx}.  In this paper we
study DM coupling to leptons only, i.e. $ l_R=e_R,\mu_R,\tau_R$, and
present all our results for the $l_R=\mu_R$ case.  The case of a
DM feebly coupled to quarks has been studied in a very similar model
in~\cite{Garny:2018ali, Garny:2018icg,Garny:2017rxs,Belanger:2018sti}.

 Considering the minimal SM extension of Eq.~(\ref{eq:lagr}), there
 are three free parameters in our model, the dark matter mass
 $m_{\chi}$, the mediator mass $m_{\sfermion}$ and  the Yukawa
 coupling $\lambda_{\chi}$. Here in particular, we use 
\begin{equation}
\label{eq:param}
	m_{\chi}, \Delta m, \lambda_{\chi}\, ,
\end{equation}
with $\Delta m=m_{\sfermion}-m_{\chi}$, as a minimal set of free
parameters of the model.  No symmetries prevent the new scalar $\phi$
to couple to the SM Higgs and this coupling will generically be
radiatively generated.  We will thus consider an extra interaction
involving the quartic coupling $ \lambda_H$:
\begin{equation}
\label{eq:Lhiggs}
	{\cal L}_{H} = -\lambda_H H^{\dag} H \sfermion^{\dag} \sfermion.
\end{equation}
We can indeed always write a box diagram involving the $4$ scalars,
through the exchanges of $Z$-boson or photon, giving rise
to~(\ref{eq:Lhiggs}), with an effective coupling $\lambda_H^{eff}\sim
g_{weak}^2/(16 \pi^2)\sim 10^{-2}$. This extra Higgs portal interaction shall
thus definitively be taken into account.  A non-negligible value of
such a coupling will add to the gauge induced $\phi$ annihilation
processes potentially giving rise to a larger $\phi$ annihilation
cross-section and modify the freeze-out of the charged scalar.  In the
rest of the paper, we will consider three representative benchmark for
this coupling, namely $\lambda_H=\{ 0.01,0.1,0.5 \}$ and we will
discuss how this impacts the viable parameter space of the model in Sec.~\ref{sec:limit-param-space}.

We would like to mention that the simplified model involving the
contributions \eqref{eq:lagr} and \eqref{eq:Lhiggs} to the Lagrangian
serves as an illustrative case to discuss in details the early
Universe dark matter production and collider prospects for a larger
class of feebly coupled dark matter scenarios with compressed mass
spectrum. The leptophilic scenario considered here can easily emerge
in non-minimal SUSY models (like extension of the MSSM) where the NLSP
is one of the right handed slepton and the DM candidate is an extra
neutralino (as for instance recently discussed in
\cite{Aboubrahim:2019qpc}).  The possibility that the NLSP is one of
right handed sleptons (not the stau) has been studied for instance in
\cite{Evans:2006sj,Calibbi:2014pza}.  Let us also emphasize that,
compared to the previous conversion driven freeze-out
studies~\cite{Garny:2018icg,Garny:2017rxs} focusing on a DM coupling
to quarks, we explicitly study the importance of the extra quartic
coupling between the mediator and the Higgs of
Eq.\eqref{eq:Lhiggs}. As mentioned above, the latter cannot be
excluded from symmetry arguments and, as we will see, can influence
the phenomenology of the model in the leptophilic case.

\section{Dark matter abundance}
\label{sec:conv}

In order to compute the number density evolution of a set of species
in kinetic equilibrium, one has to solve a coupled set of Boltzmann
equations taking the form:
\begin{eqnarray}
  H x s\frac{dY_i}{dx}&=&-\sum_{jk}\gamma_{ij\to kl}
  \left(\frac{Y_i Y_j}{Y_i^{eq} Y_j^{eq}}- \frac{Y_k Y_l}{Y_k^{eq}
    Y_l^{eq}}\right)-\sum_{jk}\gamma_{ij\to k}
  \left(\frac{Y_i Y_j}{Y_i^{eq} Y_j^{eq}}- \frac{Y_k}{Y_k^{eq}}\right)
  \label{eq:botzm0}
\end{eqnarray}
where $Y_i= n_i/s$ is the comoving number density of the species $i$,
$s$ is the entropy density, the ${eq}$ superscript refers to
equilibrium, $x=m_{ref}/T$ with $m_{ref}$  some reference mass and
$T$  the thermal bath temperature, and $H=H(x)$ is the Hubble rate
at time $x$. Here we have considered contributions from 4 particle
interactions inducing the $\gamma_{ij\to kl}$ reaction density and
contributions from 3 particle interactions inducing $\gamma_{ij\to k}$. There
is a direct correspondence between these reaction densities and the
thermal averaged scattering cross sections/decay rates going as
follows:
\begin{eqnarray}
  \gamma_{ij\to kl}&=&\int \int d\phi_i d\phi_j f_i^{eq} f_j^{eq}\int \int  d\phi_k d\phi_l (2\pi)^4\delta^4(p_i+p_j-p_k-p_l) |{\cal M}_{ij\to kl}|^2\cr
  & =&n_i^{eq}n_j^{eq}\langle \sigma_{ij\to kl}v_{ij}\rangle\\
  \gamma_{ij\to k}&=&\int \int d\phi_i d\phi_j \int  d\phi_k f_k^{eq} (2\pi)^4\delta^4(p_i+p_j-p_k) |{\cal M}_{k\to ij}|^2
   =n_k^{eq}\Gamma_{k\to ij} \frac{K_1(x)}{K_2(x)}
  \label{eq:gamm}
\end{eqnarray}
where $K_{1,2}$ denote the Bessel functions, $|{\cal M}|^2$ are the
squared scattering amplitudes summed (not averaged) over initial and
final degrees of freedom and $d\phi_i=d^3p_i/(2E_i(2\pi)^3)$.  We
  neglect quantum statistical effects and we use the Maxwell Boltzmann
  equilibrium distributions $f_i^{eq}$. See Sec.~\ref{sec:Om-lam} for
  a discussion in the context of Freeze-in.

In the standard treatment of the freeze-out~\cite{Griest:1990kh}, the
dark matter and its coannihilation partners, say $\{\chi_i\}$ with
$\chi_0$ the dark matter, are assumed to be in CE. This allows for an
important simplification of the Boltzmann equations as fast
$\chi_i\leftrightarrow\chi_j$ conversion processes imply that
$\chi_i,\chi_j$ number densities are related by
$n_i/n_i^{eq}=n_j/n_j^{eq}$. In the latter case, we recover from
Eq.~(\ref{eq:botzm0}) the familiar Boltzmann equation:
\begin{eqnarray}
 \frac{dY_{DM}}{dx}= \frac{s \langle \sigma v_{\rm eff}\rangle}{H x} \left(Y_{DM}^2-Y_{DM, eq}^2\right)
\end{eqnarray}
where $n_{DM}$ is the dark matter number density and the annihilation
cross-section is given by
\begin{eqnarray}
  \langle \sigma v_{\rm eff}\rangle&\simeq&\frac{1}{g_{\rm eff}^2}\sum_{ij} r_i r_j \langle \sigma v \rangle_{ij} \quad {\rm with } \quad g_{\rm eff}=\sum_i r_i \cr
    &{\rm and}& r_i = g_i (1+\Delta_i)^{3/2} \exp(-\, x_f \Delta_i)\,.
\label{eq:svefflim}
\end{eqnarray}
The sum in Eq.~(\ref{eq:svefflim}) runs over the co-annihilating
partners, $g_i$ is the number of internal degrees of freedom of
$\chi_i$, $\langle \sigma v \rangle_{ij}$ is the cross-section for
annihilation processes of $\chi_i\chi_j \to $~SM~SM  and
$\Delta_i=(m_i-m_0)/m_0$ where $m_0$ is the mass of the lightest of
the $\{\chi_i\}$.  Imposing $\Omega h^2=0.12$~\cite{Ade:2015xua},
assuming a dominant s-wave contribution to annihilation cross-section,
one would need $\langle\sigma v_{\rm eff}\rangle=2.2\times 10^{26}
{\rm cm}^3/s$.

\subsection {Beyond chemical equilibrium}
\label{sec:beyond-chem-equil}

In the dark matter model considered here the charged mediator $\phi$
is always in CE with the SM thermal plasma at early times because of
gauge interactions. In contrast, in e.g. the context of conversion
driven freeze-out, suppressed DM-mediator conversion processes may
prevent CE between $\chi$ and $\phi$.  The following Boltzmann system
has then to be solved (see e.g.~\cite{PhysRevD.45.455,Frigerio:2011in,
  Garny:2017rxs,Garny:2018icg}):
\begin{align}
\label{eq:BEchi}
\frac{dY_{\chi}}{dx}=\frac{-2}{Hxs} & \left[ \gamma_{\chi \chi} \left (\frac{Y^2_{\chi}}{Y^{2}_{\chi,eq}}-1\right) + \gamma_{\chi \slepton} \left(\frac{Y_{\chi} Y_{\slepton}}{Y_{\chi,eq} Y_{\slepton,eq}}-1 \right) \right. \nonumber  \\
&+  \gamma_{\chi   \to \phi }\left( \frac{Y_{\chi}}{Y_{\chi,eq}}- \frac{Y_{\slepton}}{Y_{\slepton,eq}} \right) 
+ \left. \gamma_{\chi \chi \rightarrow \slepton \slepton^{\dagger}} \left(\frac{Y^2_{\chi}}{Y^2_{\chi,eq}}- \frac{Y^2_{\slepton}}{Y^2_{\slepton,eq}}\right) \right],
\end{align}
\begin{align}
\label{eq:BEsl}
\frac{dY_{\slepton}}{dx}=\frac{-2}{ Hxs} & \left[\gamma_{\slepton \slepton^{\dagger}} \left( \frac{Y^2_{\slepton}}{Y^{2}_{\slepton,eq}}- 1\right) + \gamma_{\chi \slepton} \left(\frac{Y_{\chi} Y_{\slepton}}{Y_{\chi,eq} Y_{\slepton,eq}}-1 \right)  \right. \nonumber \\
  & -\gamma_{\chi   \to \phi }\left( \frac{Y_{\chi}}{Y_{\chi,eq}}- \frac{Y_{\slepton}}{Y_{\slepton,eq}} \right) 
- \left. \gamma_{\chi \chi \rightarrow \slepton \slepton^{\dagger}} \left(\frac{Y^2_{\chi}}{Y^2_{\chi,eq}}- \frac{Y^2_{\slepton}}{Y^2_{\slepton,eq}}\right) \right],
\end{align}
where $\gamma_{ij }=\gamma_{ij\to \alpha \beta }$, with $\alpha,\beta$
some SM particles in equilibrium with the bath, $\gamma_{\chi \to \phi
}$ includes all conversion processes, i.e. both decays and scatterings
$\gamma_{\chi \to \phi }=\left(\gamma_{\chi \alpha \to \phi
  \beta}+\gamma_{\chi \alpha \to \phi }\right)$, $Y_{\slepton}$ is the
summed contribution of {\it both} the mediator and its antiparticle
and we will use $x=m_\chi/T$.\footnote{In equation \eqref{eq:BEchi}, in
  the terms for $\chi$ self-annihilations, the factor of 2 is due to
  two $\chi$'s disappearing in each process while, in the other terms,
  it is due to the contributions both from $\phi$ and from
  $\phi^{\dagger}$. In the equation \eqref{eq:BEsl}, the factor of 2 is
  always due to the convention used here for $Y_\phi$. }  The reaction
densities have been obtained taking into account all processes given
in tables \ref{tab:coann} and \ref{tab:conv} in the Appendix
\ref{sec:processes}. We have obtained the expression of the relevant
transition amplitudes making use of {\tt FeynRules}~\cite{Alloul:2013bka} and {\tt
  Calchep}~\cite{Belyaev:2012qa}.

In general, we assume
zero initial dark matter abundance and begin to integrate our set of
equations at $x=0.01$ i.e., in the small mass splitting limit, a
time at which we expect the mediator is relativistic and in
equilibrium with the SM bath. Also notice that the above set of
equations assume kinetic equilibrium between the dark matter and the
mediator, we comment further on this assumption in the next section.

\subsection{DM  abundance dependence on the conversion parameter}
\label{sec:relic}

\begin{figure}
  \begin{center}
  \includegraphics[scale=0.4]{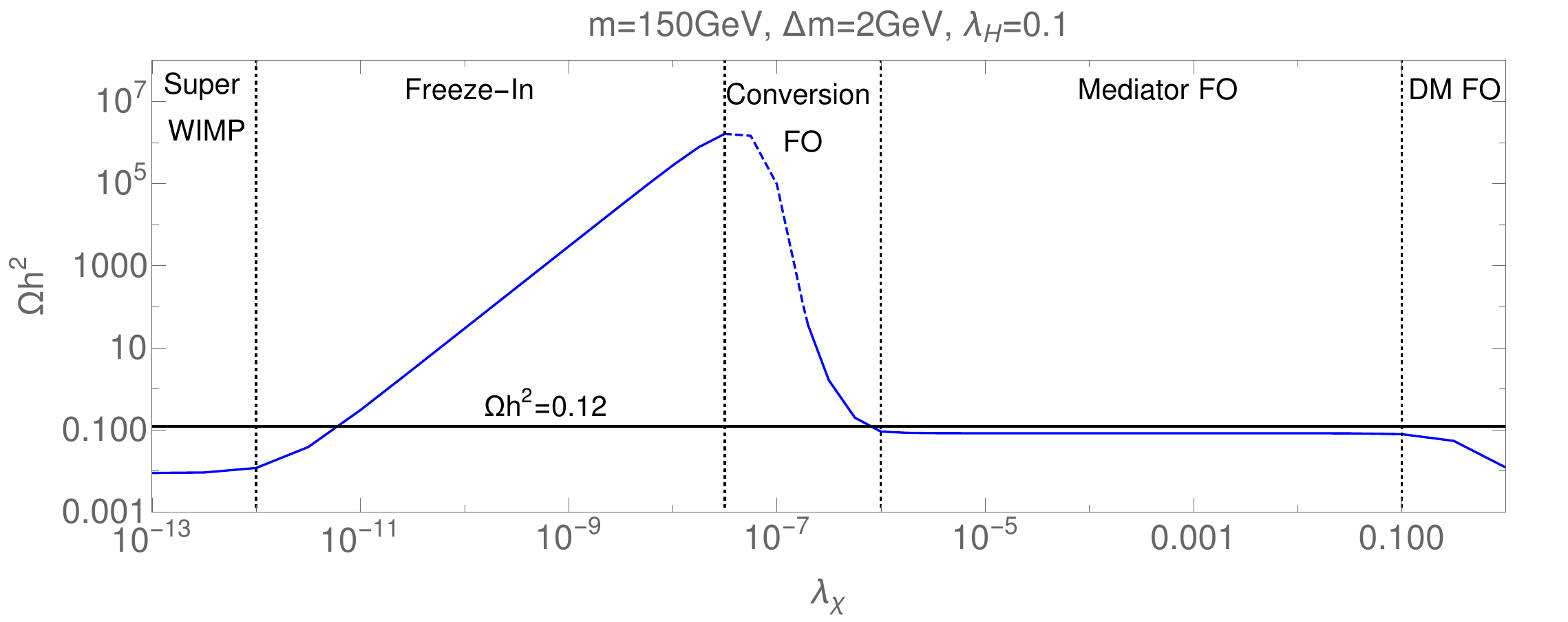}  
  \end{center}
  \caption{DM abundance as a function of the Yukawa coupling for
    $m_\chi=150$ GeV, $\Delta m=2$ GeV and $\lambda_H=0.1$ (i.e. for
    compressed DM-mediator mass spectrum).}
  \label{fig:ann-conv}
\end{figure}

 Figure~\ref{fig:ann-conv} shows the dependence of the DM relic
 abundance with respect to the Yukawa coupling $\lambda_\chi$ in the
 case of compressed mass spectrum. For illustration, we consider a
 coupling to $l_R=\mu_R$ with $m_{\chi} = 150$ GeV, $\Delta m = 2$ GeV
 and $\lambda_{H} = 0.1$, and we comment on the different regimes of
 DM production that are realized by varying $\lambda_{\chi}$.
 
 For the largest values of the coupling, that is $\lambda_\chi > 0.1$,
 the relic abundance decreases with increasing coupling. This is the
 well known ``standard WIMP behavior'' with $\Omega h^2\propto
 1/\langle \sigma v\rangle_{\chi\chi}$ where $\langle \sigma
 v\rangle_{\chi\chi}$ denotes the cross-section for the dominant
 annihilation processes ($\chi \chi \to$~SM~SM) with $\langle \sigma
 v\rangle_{\chi\chi}\propto \lambda_\chi^4$. In the latter case, the
 freeze-out is {\it DM annihilation driven} meaning that freeze-out
 occurs when the rate of DM annihilation, initially maintaining DM in
 equilibrium with the SM bath, becomes smaller than the Hubble
 rate. More generally, in the standard freeze-out regime (DM in CE),
 the DM abundance goes as $\Omega h^2\propto 1/\langle \sigma
 v\rangle_{eff}$, with $\langle \sigma v\rangle_{eff}$ defined in
 Eq.~(\ref{eq:svefflim}).  Around $\lambda_\chi\sim 0.1$, the $\Omega
 h^2$ curve bends and we enter in the {\it co-annihilation driven} regime. In the
 latter case $\chi \phi\to$ SM SM processes can play the main role
 with $\langle \sigma v\rangle_{eff}\propto \langle \sigma
 v\rangle_{\phi\chi} \exp(- x_f\Delta m/m_\chi)$ where $x_f\sim 30$ and
 $\langle \sigma v\rangle_{\phi\chi}\propto \lambda_\chi^2$. For lower
 values of the coupling, $10^{-6}\lesssim\lambda_\chi\lesssim 0.1$,
 the relic abundance eventually becomes driven by the rate of mediator
 annihilations, i.e. $\phi \phi^\dag \to$~SM~SM, with $\langle \sigma
 v\rangle_{eff}\propto \langle \sigma v\rangle_{\phi\phi^\dag} \exp(-2
 x_f\Delta m/m_\chi)$. The cross-section $\langle\sigma
 v\rangle_{\phi\phi^\dag}$ mainly depends on the gauge and quartic
 couplings $g, \lambda_H \gg \lambda_\chi$ and the DM relic density
 becomes thus independent of the Yukawa coupling $\lambda_\chi$.  The
 blue curve of Fig.~\ref{fig:ann-conv} turn to a constant and the
 freeze-out has become {\it mediator annihilation driven}. The
 parameter space for annihilating/co-annihilating WIMP model of
 Eq.~(\ref{eq:lagr}) has already been studied in details, see
 e.g.~\cite{Garny:2015wea} and references therein.

 If chemical equilibrium between the DM and the other species could be
 maintained for even lower values of $\lambda_\chi$ we would expect
 the relic abundance to stay independent of this
 parameter. Refs.~\cite{Garny:2017rxs} however pointed out that a new
 window for dark matter production opens at sufficiently small values
 of $\lambda_\chi$. For suppressed rate of conversion processes, here
 with $\lambda_\chi \lesssim 10^{-6}$, we are left with a dark matter
 abundance that is larger than in the equilibrium case due to
 inefficient $\chi\to\phi$ conversions. We enter in the {\it
   conversion driven freeze-out} regime with a relic abundance that is
 larger than in the DM or mediator annihilation driven regime and that
 increases again with decreasing $\lambda_\chi$.  This behavior is due
 to the rate of conversion processes becoming slower than the Hubble
 rate.  In contrast, in the ``standard WIMP'' freeze-out, it is the
 rate of the relevant (co-)annihilation processes that become
 inefficient.

 As it is well known for even lower value of the coupling,
 $\lambda_\chi < 10^{-8}$, the dark matter is expected to freeze-in,
 see e.g.~\cite{Hall:2009bx}. The dark matter is produced through
 decays or scatterings of the mediator, i.e. $\Omega h^2\propto
 \lambda^2_\chi$, but the suppressed coupling $\lambda_\chi$ prevents
 $\chi$ to reach its equilibrium density before it freezes-in. In the
 latter case, contrarily to the freeze-out, the relic abundance
 eventually decreases thus with decreasing values of the
 coupling. Notice that in the freeze-in regime inverse
 decay/scattering processes giving rise to $\chi\to\phi$ can be
 neglected, see e.g.~\cite{Hall:2009bx,Belanger:2018ccd}.\footnote{
   Setting the scatterings (the decays) to zero by hand, we recover
   the results presented in~\cite{Hall:2009bx} assuming that {\it
     only} the mediator is in kinetic equilibrium with the thermal
   bath and produces the dark matter through decays (scatterings).}
 Finally, for couplings $\lambda_\chi<10^{-12}$, the relic abundance
 gets eventually settled through the superWIMP mechanism,
 see~\cite{Covi:1999ty,Feng:2003uy,Garny:2018ali}. In the latter case,
 the DM is produced through the decay of the thermally decoupled
 mediator (after $\phi$ freezes-out) and $\Omega_\chi
 h^2=m_\chi/m_\phi \,\Omega_\phi h^2$. In the framework considered
 here $m_\chi/m_\phi\approx 1$ and the DM relic abundance is just
 fixed by the mediator freeze-out. As a result the abundance curve
 becomes again independent of the conversion factor $\lambda_\chi$ and
 we have $\Omega_\chi h^2\approx\Omega_\phi h^2$, ie effectively
 $\langle \sigma v\rangle_{eff}\approx\langle \sigma
 v\rangle_{\phi\phi^\dag}$. This regime looks a priori very similar to
 the mediator annihilation driven freeze-out case. The plateaus in the
 superWIMP regime ($\lambda_\chi<10^{-12}$) and in the mediator
 annihilation driven freeze-out regime
 ($10^{-6}\lesssim\lambda_\chi\lesssim 0.1$) correspond however to two
 different values of $\Omega_\chi h^2$. This is because in the latter
 case the relevant effective annihilation driving the relic abundance
 $\langle \sigma v\rangle_{eff}$ gets an extra mass splitting
 dependent Boltzmann factor ($\exp(-2 x_f\Delta m/m_\chi)$) and both
 the mediator and the DM degrees of freedom have to be taken into
 account in its computation, see eq.~(\ref{eq:svefflim}).

 With the dashed line in Fig.~\ref{fig:ann-conv} we show the ``naive''
 transition between the conversion driven freeze-out and freeze-in
 regime that one would obtain using Eqs.~(\ref{eq:BEchi})
 and~(\ref{eq:BEsl}).  In this picture, one can compute $\phi$ and
 $\chi$ abundances assuming that they stay in kinetic equilibrium all
 the way from standard freeze-out to freeze-in. This is difficult to
 argue. In Ref.~\cite{Garny:2017rxs}, the authors compared the results
 of the un-integrated Boltzmann equations with the one of
 Eqs.~(\ref{eq:BEchi}) and~(\ref{eq:BEsl}). It was shown that even
 though kinetic equilibrium can not be maintained all along the
 process of conversion driven freeze-out, the resulting error on the
 estimation of the relic dark matter was small ($\sim 10\%$). The
 authors argue that this is due to the thermally coupled mediator
 eventually decaying to the DM at the time of DM freeze-out actually
 allowing for the dark matter to inherit back a thermal distribution
 (or at least close enough to it). For this effect to be relevant, the
 DM abundance should definitively not be too far higher than the
 mediator abundance around freeze-out.  In Fig.~\ref{fig:ann-conv},
 the dashed line starts at $\lambda_\chi=2 \cdot 10^{-7}$ and denotes the cases when
 $Y_\chi- Y_\phi>0.1 \times Y_\chi$ around the time the abundance of $\chi$ freezes. In the
 latter case, we assume that we are in the same situation as
 in~\cite{Garny:2017rxs} and we neglect departure from kinetic
 equilibrium. We have checked that all viable models considered in the
 following for conversion driven freeze-out satisfy to this 10\%
 condition.

  In what follows we study in more details two benchmark points of
  Fig.~\ref{fig:ann-conv} giving rise to $\Omega h^2=0.12$ in the
  regime of conversion driven freeze-out and freeze-in (with a
  coupling to the SM muon and $m_{\chi} = 150$ GeV, $\Delta m = 2$
  GeV, and $\lambda_{H} = 0.1$).  For the latter purpose, we compare,
  for fixed value of $\lambda_\chi$, the rates of interaction involved
  in Eqs.~(\ref{eq:BEchi}) and (\ref{eq:BEsl}) to the Hubble rate in
  Figs~\ref{fig:Rateplot3} and \ref{fig:Rateplot1} and show the
  associated evolution of the DM and mediator yields in
  Figs.~\ref{fig:Yield3} and ~\ref{fig:Yield1}. The plotted rates
  $\Gamma$ for a given process in Figs.~\ref{fig:Rateplot3}
  and~\ref{fig:Rateplot1} are taken to be $\Gamma_{ij\to k(l)}=
  \gamma_{ij\to k(l)}/n^{eq}_{\chi}$, except for the rate of $\phi$
  annihilation in which case $\Gamma_{\phi\phi^\dag\to {\rm SM\,SM}}=
  \gamma_{\phi\phi^\dag}/n^{eq}_{\phi}$. 
  In
  addition let us emphasize that contrarily to the annihilation driven
  freeze-out, the DM abundance through freeze-in depends on the
  initial conditions assumed for $Y_\chi$, see
  e.g.~\cite{Hall:2009bx}. For conversion driven freeze-out there
  can be a dependence on the initial conditions as well, see
  Appendix~\ref{sec:InitCond} for a discussion. All along this paper
  we assume a negligible initial $\chi$ abundance.

\subsubsection{Conversion driven freeze-out}
\label{sec:illustr-benchm-model}

\begin{figure}
  \centering
  \subfloat[]{\label{fig:Rateplot3}{\includegraphics[scale=0.54]{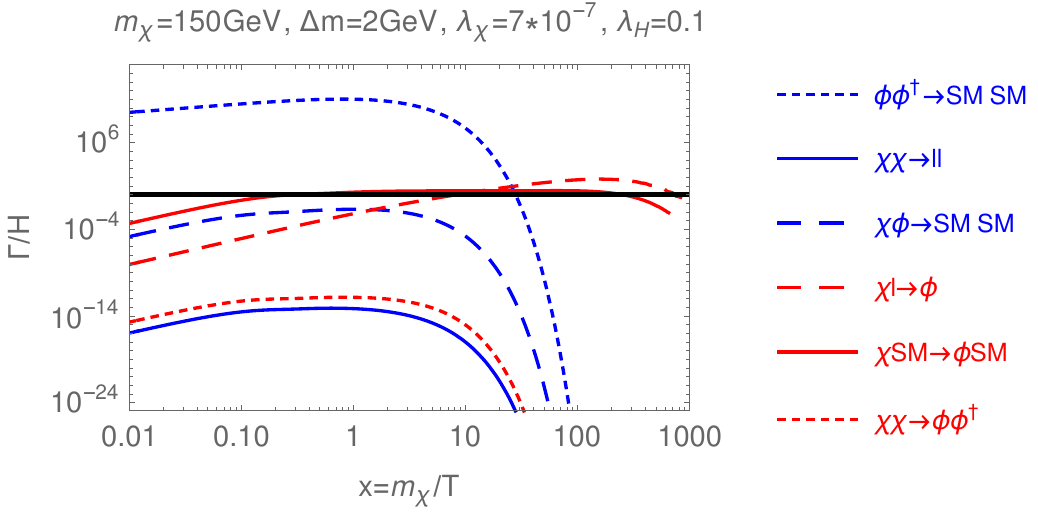}}}
  ~~
  \subfloat[]{\label{fig:Yield3}{\includegraphics[scale=0.54]{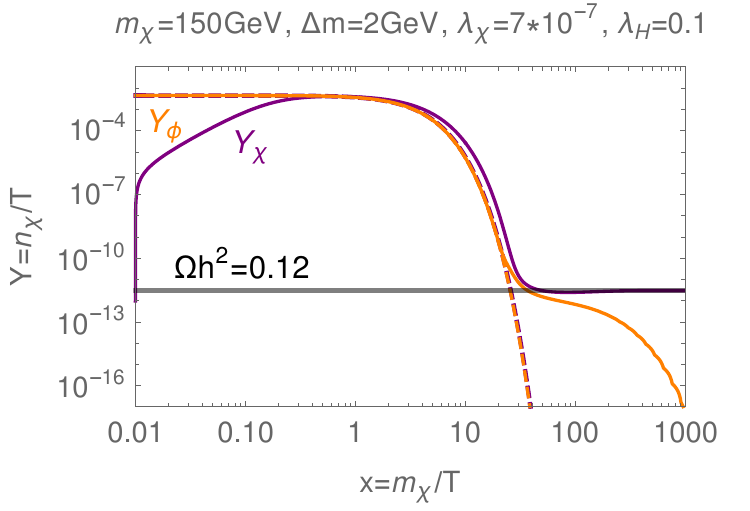}}}\\
  \subfloat[]{\label{fig:Rateplot1}{\includegraphics[scale=0.54]{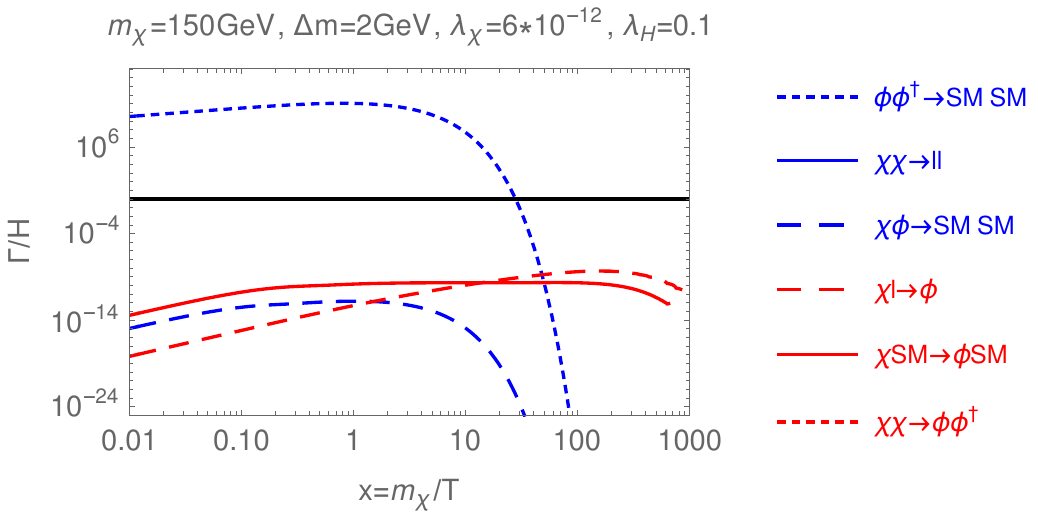}}}
   ~~
   \subfloat[]{\label{fig:Yield1}{\includegraphics[scale=0.54]{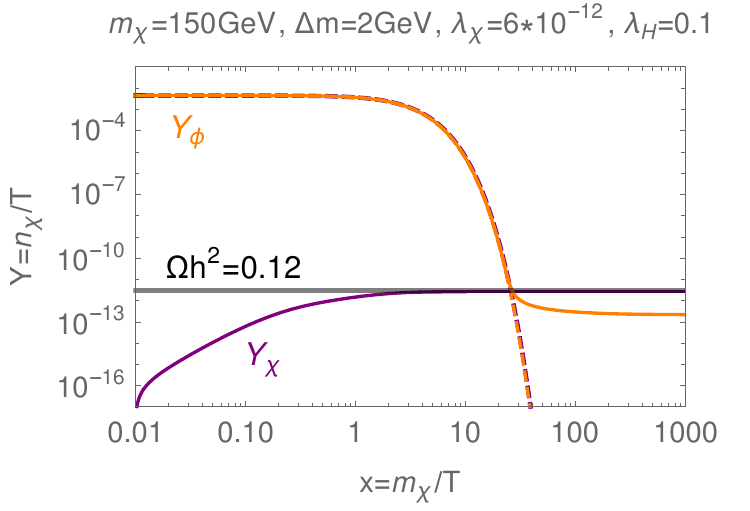}}}
  \caption{Benchmark points of Fig.~\ref{fig:ann-conv} giving rise to
    $\Omega h^{2}=0.12$. Left: Ratio of the rates of interactions
    depicted in the legend and of the Hubble rate as a function of
    $x=\frac{m_{\chi}}{T}$. The conversion processes are depicted in
    red, the (co-)annihilation ones in blue. The black line depicts
    $\Gamma/H=1$. Right: Evolution of the yield of $\chi$ (solid
    purple line) and $\phi$ (solid orange line) and their equilibrium
    yield (dashed lines) as a function of $x$.  Notice that, given the
    small $\Delta m$ the two equilibrium curves (dashed) are almost
    indistinguishable.}
    \label{fig:EvolutionComp}
\end{figure}

Within the set up considered in Fig.~\ref{fig:ann-conv}, we obtain
$\Omega h^2=0.12$ through conversion driven freeze-out for
$\lambda_{\chi} = 7 \cdot 10^{-7}$.  The associated rate and
abundance evolution are shown in Figs.~\ref{fig:Rateplot3}
and~\ref{fig:Yield3}.  We readily see in Fig.~\ref{fig:Yield3} that
the $\chi$ population does not follow its equilibrium number density
evolution and that before the standard freeze-out temperature (or
equivalently $x= m_\chi/T \sim 30$) it gives rise to $n_\chi (T) >
n_\chi^{eq} (T)$ due to the suppressed conversion processes that are unable
remove the overabundant $\chi$ efficiently.

 Considering Fig.~\ref{fig:Rateplot3}, we confirm that we are in a
 regime in which the most efficient process in reducing the dark
 sector abundance is the mediator annihilation (short dashed blue
 line). In contrast, coannihilation (long dashed blue) and especially
 annihilation processes (continuous blue and short-dashed red) are
 always suppressed. On the other hand some of the conversion rates can
 be of the order of, or faster than, the Hubble rate before the
 mediator freezes-out, as can be seen with the continuous and
 long-dashed red lines. Those processes will thus affect both the DM
 and the mediator abundance along the time. For the benchmark model
 considered here the 4 particles interactions ($\chi $SM$ \to\phi $SM)
 dominate at early times while the 3 particles interaction (inverse
 mediator decay) play the leading role around the freeze-out
 time. Given that we have considered $Y_{\chi}=0$ initially, the
 scatterings $\chi $SM$ \to\phi $SM appear to play the most important
 role in bringing $Y_{\chi}$ near to its equilibrium value at early
 time.  In Fig.~\ref{fig:Yield3}, we see that DM yield always deviates
 from its equilibrium value. It is lower than the equilibrium yield at
 early time and larger than the equilibrium yield when approaching the
 freeze-out time. It is also noticeable that when
 $Y_{\chi}>Y_{\chi}^{eq}$, the DM yield still stays close the its
 equilibrium yield until the mediator decouples from the thermal
 bath. This is due to barely but still efficient conversion
 processes. This behavior has already been described in length
 in~\cite{Garny:2017rxs}. We discuss further the viable parameter
 space and detection prospects in the leptophilic scenario in
 Sec.~\ref{sec:viab}.

\subsubsection{Freeze-in from mediator decay \emph{and} scatterings. }
\label{sec:Om-lam}

 One can also account for all the DM through the freeze-in
 mechanism for the lowest values of the conversion coupling in
 Fig.~\ref{fig:ann-conv}, that is $\lambda_\chi= 6\cdot 10^{-12}$.
 The associated rates and abundance evolution are shown in
 Fig.~\ref{fig:Rateplot1} and~\ref{fig:Yield1}. In
 Fig.~\ref{fig:Yield1} we mainly recover the standard picture of the
 dark matter freeze-in which relic abundance is due to scatterings and
 decays of a mediator in thermodynamic equilibrium with the SM
 bath. The relic dark matter abundance freezes-in around the time at
 which the rate of mediator decay/scatterings gets strongly suppressed ($x\sim
 3$).  Notice that after the mediator freezes-out its relic population
 eventually decay to DM and leptons at a time characterized by its
 life-time $\tau_\phi$. This would correspond to the ``superWIMP''
 contribution to the relic abundance. For the benchmark considered
 here $\tau_\phi= 9s$ which is safely orders of magnitudes below BBN
 bounds\footnote{An a approximate BBN bound can be estimated from the
   analysis presented in Ref.~\cite{Jedamzik:2006xz} for
   electromagnetic decays. Considering that in Fig.~\ref{fig:Yield1}
   the relic abundance $\Omega_\phi h^2$ of $\phi$ before its decay is
  one order of magnitude lower than $\Omega h^2=0.12$,
   we expect a BBN bound of $\tau_\phi\lesssim 10^6$ s.}  and the
 superWIMP contribution to the DM relic abundance is around one order of 
 magnitude lower than the freeze-in one. In the rest
 of this section, we restrict our discussion to the freeze-in
 contribution to the dark matter relic abundance while neglecting the superWIMP
 contribution. Note that in Fig.~\ref{fig:ann-conv} and following sections, the superWIMP contributions 
 have been taken into account.
 For a detailed study on the freeze-in and superWIMP interplay
 see e.g.~\cite{Garny:2018ali}.

For reference, we remind that the simplest freeze-in model involves
a mother particle $A$ in thermal equilibrium with the SM that decays
to a bath particle $B$ and to DM $\chi$.  In such a scenario the DM
comoving number density induced through freeze-in is  associated to the
{\it decay only} of $A \to B\,\chi$, and
the predicted dark matter density simply reduces
to~\cite{Hall:2009bx}:
\begin{equation}
\label{FI_Hall}
Y_{\chi} = \frac{135 g_A}{ 1.66 \times 8 \pi^3 g_*^{3/2}} \frac{M_{Pl} \Gamma_A}{m_A^2}\,,
\end{equation}
where $g_A$ counts the spin degrees of freedom of the mother particle
$A$, $g_*$ is the number of degrees of freedom at the freeze-in
temperature $T \sim m_A$, and $M_{Pl}= 1.22 \cdot 10^{19}$ GeV is the
Planck mass. This result is however obtained neglecting all potential
contributions from scatterings at any time since they are typically
sub-dominant compared to the decays~\cite{Hall:2009bx}.  In the case
considered here, the equation (\ref{FI_Hall}), with the mother
particle $A\equiv \phi$, underestimates the relic dark matter
abundance in the freeze-in regime.  One reason for this is the small
mass splitting between the mediator and the DM implying that the decay
of the mediator is kinematically suppressed.  The contribution to the
DM abundance from scattering processes hence becomes more relevant,
and this feature is not captured by the simple expression of
Eq.~\eqref{FI_Hall}, see also the discussion
in~\cite{Belanger:2018ccd}.

\begin{figure}
  \begin{center}
  \includegraphics[scale=0.65]{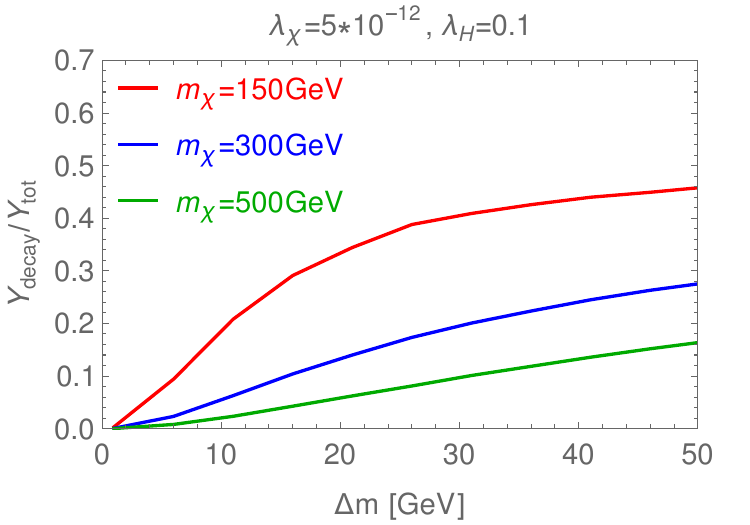}  
  \end{center}
  \caption{The ratio of the freeze-in yields $Y^{fi}_{\rm decay}/Y^{fi}_{\rm tot}$ as a function of $\Delta m$
  when taking only decay processes into account for  $Y^{fi}_{\rm decay}$ and including all
 the relevant scattering and decay processes for $Y_{\rm tot}^{fi}$. We show the results for three different values of the DM mass $m_\chi \in \{150, 300, 500 \}$ GeV while keeping $\lambda_{\chi} = 5 \cdot 10^{-12}$ and $\lambda_H=0.1$ fixed.}
  \label{fig:FI_ratio}
\end{figure}

 The relative importance of the decay contribution to the freeze-in is
 displayed as a function of the mass splitting in
 Fig.~\ref{fig:FI_ratio}. The red curve correspond to an illustrative
 benchmark with $m_\chi= 150$ GeV, $\lambda_\chi= 5 \cdot 10^{-12}$
 and $\lambda_H=0.1$.  We show the ratio between the DM abundances
 obtained through freeze-in $Y^{fi}_{\rm decay}/Y^{fi}_{\rm tot}$
 considering the decay processes only for $Y^{fi}_{\rm decay}$
 (setting the scattering processes to zero by hand) and including all
 the relevant scattering and decay processes for $Y_{\rm
   tot}^{fi}$. For small mass splittings, the mediator decay
 contribution is subleading.  For increasing mass splitting, instead,
 the decay process becomes the main player.  Let us emphasize though
 that even for the largest values of the mass splitting considered in
 Fig.~\ref{fig:FI_ratio} the contribution from scattering is non
 negligible.  This is due to the fact that, in the model under study,
 a large number of scattering processes can contribute to the DM
 production. There are indeed $\sim $ 10 possible scattering processes
 with a rate of DM production $\propto \lambda_\chi^2\alpha$ to be
 compared to one single decay process with a rate $\propto
 \lambda_\chi^2$, where $\alpha$ is the fine structure constant
 (Tab.~\ref{tab:conv} in Appendix \ref{sec:processes} lists all the relevant processes).  The
 multiplicity factor of the scattering process partially compensates
 for the extra SM gauge coupling suppression $\alpha$. As a result,
 scattering contribution to the dark matter relic abundance through
 freeze-in can still be $\sim {\cal O} (1)$ compared to the decays for sizeable mass
 splitting. The relative contribution of the decay is also a function
 of the mediator mass. The larger the mediator mass, the smaller is
 the decay contribution at fixed value of the couplings and of the
 mass spitting (see the blue and green curves in
 Fig.~\ref{fig:FI_ratio}). This is because the decay rate scales like
 $\Gamma_{\phi} \sim m_{\phi}^{-1}$, see Sec.\ref{sec:exp}.

 The fact that scatterings can dominate the freeze-in production can
 also be seen in Fig.~\ref{fig:Rateplot1}, in the case of a freeze-in
 benchmark scenario with small mass splitting. We indeed see that the
 scattering processes are more efficient than the decay process for
 the entire period in which the mediator abundance is
 unsuppressed. 
We conclude
that both scatterings and decay contribution must be taken into
 account here to provide a correct estimate of the relic dark matter
 abundance through freeze-in.\footnote{We have checked that the
   freeze-in is still dominated by IR physics as integrating the
   equations from any point with $x<0.01$ (i.e. a temperature high
   enough compared to the mediator mass) does not change the
   behavior.}

Let us also mention that throughout this paper, we neglect quantum
statistical effects and we use the Maxwell Boltzmann equilibrium
distributions $f_i^{eq}$. Notice however that, when considering the DM
production mechanisms for suppressed dark matter couplings and small
mass, splittings between the DM and the mediator the precise
statistics can be relevant. Taking into account Fermi Dirac statistics
for the lepton and DM produced by freeze-in through decay can actually
give rise up to a 50\% suppression of the DM abundance, see the {\tt
  micrOMEGAs5.0} paper \cite{Belanger:2018ccd} for a
discussion. Making use of {\tt micrOMEGAs5.0}, we have explicitly
checked, that for the benchmark model of Fig.~\ref{fig:ann-conv} the
suppression is always $\sim$ 50\% in the freeze-in region while, for
the typical mass range and mass splittings considered here, the
suppression is always between 10 and 50\%. Overall we do not expect
this effect to qualitatively change the results presented here.

\section{Phenomenology of conversion driven freeze-out}
\label{sec:viab}

\begin{figure}
  \centering
  \subfloat[$\lambda_H = 0.01$]{\label{fig:ParamPlot1e-2}
    \includegraphics[scale=0.6]{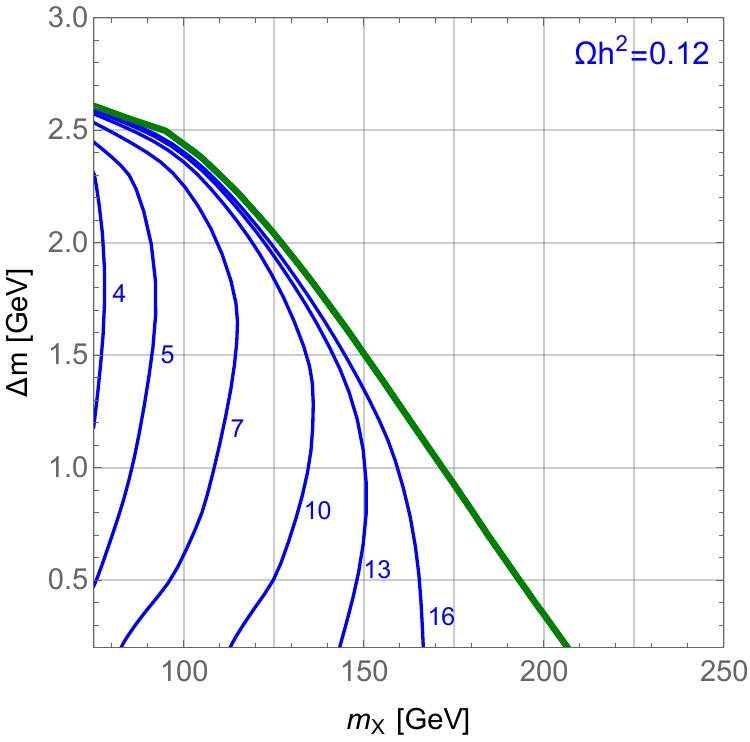}}
    \subfloat[$\lambda_H = 0.1$]{\label{fig:ParamPlot1e-1}
    \includegraphics[scale=0.6]{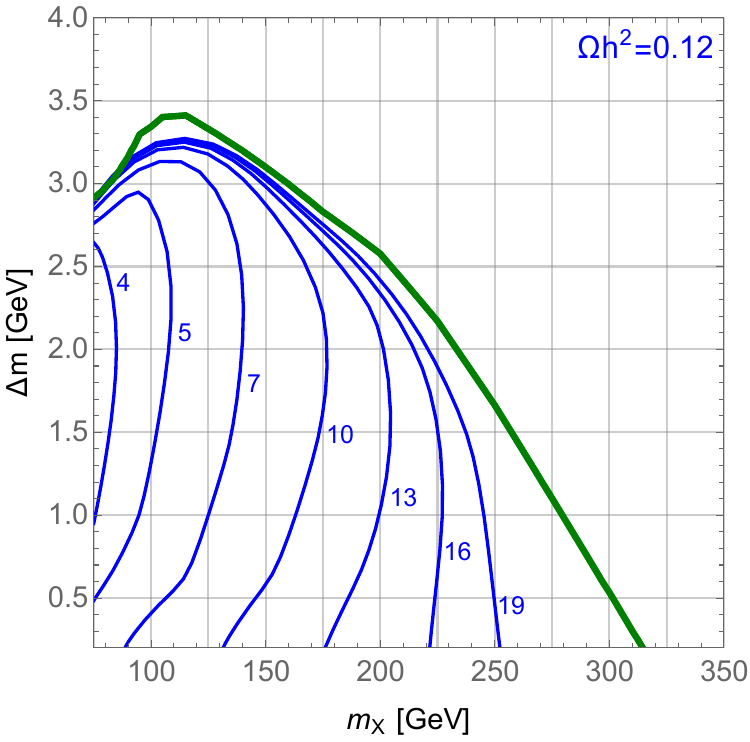}}
  \\
  \subfloat[$\lambda_H = 0.5$]{\label{fig:ParamPlot5e-1}
    \includegraphics[scale=0.6]{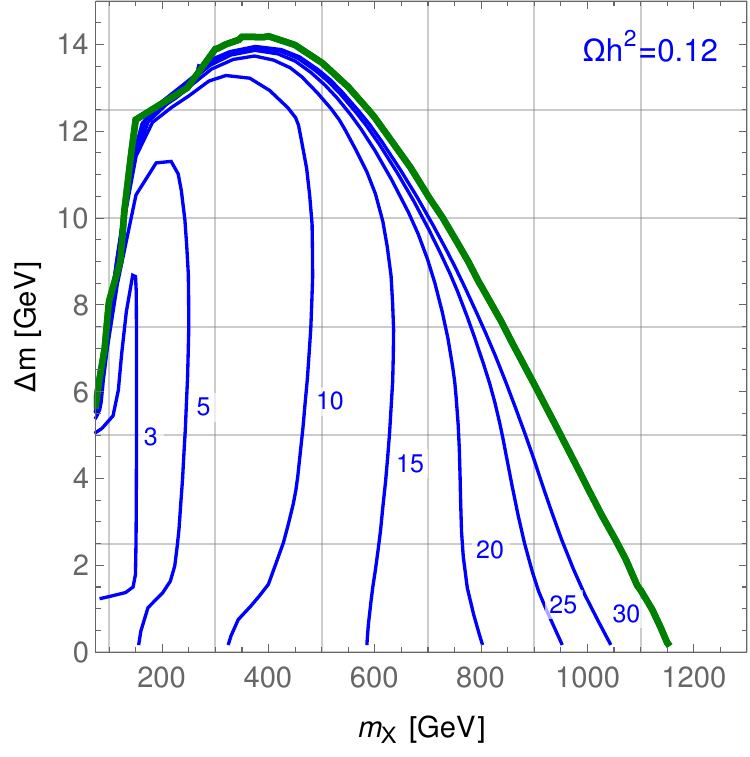}}
    \caption{Viable parameter space for DM abundance through
      conversion driven freeze-out for several value of the $H-\phi$
      coupling $\lambda_H$. Contours denoting $\Omega h^2 = 0.12$ for fixed value of the Yukawa
      coupling $\lambda_\chi/10^{-7}$ are shown with blue lines. The border of
      the parameter space is delimited by a green contour
      corresponding the combinations of $\Delta m, m$ giving rise to
      the right dark matter abundance through mediator annihilation
      driven freeze-out. }
  \label{fig:ParamPlot}
\end{figure}

We focus now on the new region of the parameter space with feeble
couplings giving rise to all the DM through the conversion driven
freeze-out. This DM production regime opens up for compressed
mediator-DM mass spectrum, and suppressed conversion couplings such
that the DM is out of CE, as described in Sec.~\ref{sec:illustr-benchm-model}.
Notice that the DM freeze-in has been considered in a similar
  scenario in Ref.~\cite{Belanger:2018sti} while the (co-)annihilation
  freeze-out has been considered in
  ~\cite{Garny:2011ii,Garny:2015wea,Garny:2013ama,Bringmann:2012vr,Kopp:2014tsa,
    Giacchino:2013bta,Giacchino:2014moa,Giacchino:2015hvk,
    Colucci:2018vxz,Toma:2013bka,Ibarra:2014qma}.  
    We discuss the
viable parameter space (i.e. $\Omega h^2=0.12$) for conversion driven
freeze-out in Sec.~\ref{sec:limit-param-space} and we study the
collider constraints in Sec.~\ref{sec:exp}.

\subsection{Viable parameter space}
\label{sec:limit-param-space}

In the 3 plots of Fig.~\ref{fig:ParamPlot}, corresponding to 3 values
of the Higgs-mediator coupling $\lambda_H$, the viable parameter space
 for conversion driven freeze-out is enclosed
by the green line in the plane $(m_\chi,\Delta m)$. We readily see
that larger values of $\lambda_H$ give rise to a larger viable
parameter space.  Let us first comment on the area above the green
contour. In the latter region, the standard freeze-out mechanism
(with DM in CE) can give rise to all the DM for a specific choice of
the coupling $\lambda_{\chi}$. For large $\Delta m$ or $m_\chi$ (above
the green line), the freeze-out is {\it DM annihilation driven}, i.e. the $\langle \sigma v\rangle_{eff}$ of
Eq.~(\ref{eq:svefflim}) is directly given by the dark matter
annihilation cross-section $\langle\sigma
v\rangle_{\chi\chi}$. Approaching the green line at fixed value of
$m_\chi$ implies smaller mass splitting $\Delta m$ and thus a larger
contribution of co-annihilation processes to $\langle \sigma
v\rangle_{eff}$. The green line itself delimiting the viable parameter
space for conversion driven freeze-out is obtained by requiring that
{\it mediator annihilation driven} freeze-out give rise to all the DM
(DM still in CE). In the latter case, $\langle \sigma v\rangle_{eff}$
is directly proportional to the mediator annihilation cross section $
\langle \sigma v\rangle_{\phi\phi^\dag}$ and to the Boltzmann
suppression factor $ \exp(-2 x_f\Delta m/m_\chi)$ with $x_f\approx
30$. Let us emphasize that the green border, in the 2-dimensional
$(m_\chi,\Delta m)$ plane, can be realized for a wide range of suppressed
conversion couplings. In e.g. the specific benchmark case of
Fig.~\ref{fig:ann-conv}, the mediator driven freeze-out region
extended from $\lambda_{\chi} \sim 10^{-1}$ to $\lambda_{\chi} \sim
10^{-6}$ giving rise to a fixed value of $\Omega h^2$. Notice that the
entire region of the parameter space above the green line has already
been analyzed in details in previous studies, see
e.g.~\cite{Garny:2015wea}, and we do not further comment on this
region here.

Below the green line, the standard freeze-out computation (assuming DM
in CE) would predict an underabundant DM population. The conversion
coupling gets however so suppressed that the DM can not any more be
considered in CE and the standard computation involving
Eq.~(\ref{eq:svefflim}) breaks down. In contrast, the treatment of the
Boltzmann equations described in Sec.~\ref{sec:beyond-chem-equil} can
properly follow the DM yield evolution in such a region. As a result
the blue contours in Fig.~\ref{fig:ann-conv} can give rise to $\Omega
h^2=0.12$ through {\it conversion driven} freeze-out (DM out of CE)
for fixed value of $\lambda_\chi \in$ few $\times [10^{-7},
  10^{-6}]$.

We also notice that the maximum value of the allowed mass splitting
and of the dark matter mass in the conversion driven region increase
with the Higgs portal coupling, going from $\Delta m_{max}\simeq 2.6 $
GeV and $ m_\chi^{max}=180$ GeV for the minimal value of $\lambda_H=
0.01$ to $\Delta m_{max}\simeq 14 $ GeV and $ m_\chi^{max}=1$ TeV for
e.g.~$\lambda_H= 0.5$. Increasing $\lambda_H$ effectively increases
the resulting $\langle \sigma v\rangle_{\rm eff}$ and, as a consequence, decreases the
DM relic abundance through mediator annihilation driven freeze-out which is relevant for the
extraction of the green contour. In
order to understand this behavior let us comment on the overall shape
of the viable parameter space delimited by the green line.  This line
corresponds to models giving $\Omega h^2=0.12$ considering mediator
annihilation driven freeze-out, i.e. $\langle \sigma
v\rangle_{eff}\propto \langle \sigma v\rangle_{\phi\phi^\dag} \exp(-2
x_f\Delta m/m_\chi)$ where $\langle\sigma v\rangle_{\phi\phi^\dag}$
mainly depends on $\alpha, \lambda_H$.  It is easy to check that the
general dependence of the countour (hill shape) just results from the
competition between the two factors $\langle \sigma
v\rangle_{\phi\phi^\dag}$ and $\exp(-2x_f\Delta m/m_\chi)$.

Let us first focus on the large mass range, where the green line
display a negative slope in the $(m_{\chi}, \Delta m)$ plane.  At
fixed value of $m_\chi$, decreasing $\Delta m$, one decreases $\Omega
h^2$.  This is due to the $\propto \exp(-2x\Delta m/m_\chi)$
dependence of the annihilation cross-section. One way to compensate
for this effect and anyway obtain the correct relic abundance is to
consider larger values of the mediator mass (or equivalently the DM
mass) as one can expect $\langle \sigma v\rangle_{\phi\phi^\dag}
\propto m_\phi^{-2}$ for large enough $m_\phi$.  As a result, for
fixed $\Omega h^2$, lower $\Delta m$ implies larger $m_\phi$. This is
indeed the shape of the green line that we recover for large $m_\phi$
in the plots of Fig.~\ref{fig:ParamPlot}. At some point $\Delta m<m_l$
where $l$ is the SM fermion involved in the Yukawa
interaction~(\ref{eq:lagr}) and we get to the maximum allowed value of
$m_\phi$ (focusing on 2 body decay $\phi \to \chi l$ only\footnote{For an analysis involving 3 body decays, see~\cite{Khoze:2017ixx}.}). By the
same token, one simple way to enlarge the parameter space allowing for
larger $m_\chi$ at fixed $\Delta m$ is to increase the annihilation
rate of the mediator, which can be done by increasing $\lambda_H$. 
Therefore, increasing $\lambda_H$ implies a larger viable value of $m_\phi$ 
(or equivalently $m_\chi$) to
account for the right DM abundance. Going back to the benchmark of
Fig.~\ref{fig:ann-conv}, this would imply the horizontal part of the
$\Omega h^2$ curve goes to lower $\Omega h^2$ value when increasing
$\lambda_H$. This is shown with the red curve of
Fig.~\ref{fig:multiLambdaPlot} in appendix~\ref{sec:processes}. Also,
as illustrated in Fig.~\ref{fig:multiLambdaPlot}, this effect can be
compensated either by increasing the DM/mediator mass (orange curve in Fig.~\ref{fig:multiLambdaPlot})
or by increasing the mass splitting (green curve in Fig.~\ref{fig:multiLambdaPlot}) as expected from the
green contours of Fig.~\ref{fig:ParamPlot}.

Finally, we comment on the shape of the green line in the small mass
region, where it has a positive slope in the $(m_{\chi}, \Delta m)$
plane. This time this is mainly due to the $\exp(-2 x_f\Delta
m/m_\chi)$ factor in $\langle \sigma v\rangle_{eff}.$ In the small
mass region the exponential suppression is dominant in determining the
shape.  Larger $\Delta m$ hence requires larger $m_{\chi}$ in order to
keep the dark matter abundance at the correct value.  Note that this
region is present also in the small $\lambda_{H}$ case, but it is
realized for dark matter masses lower than the ones showed in the
plots (and not phenomenologically interesting, see next section).

\subsection{Collider constraints }
\label{sec:exp}

\begin{figure}
	\centering
	\subfloat[$\lambda_H = 0.01$]{\includegraphics[scale=0.6]{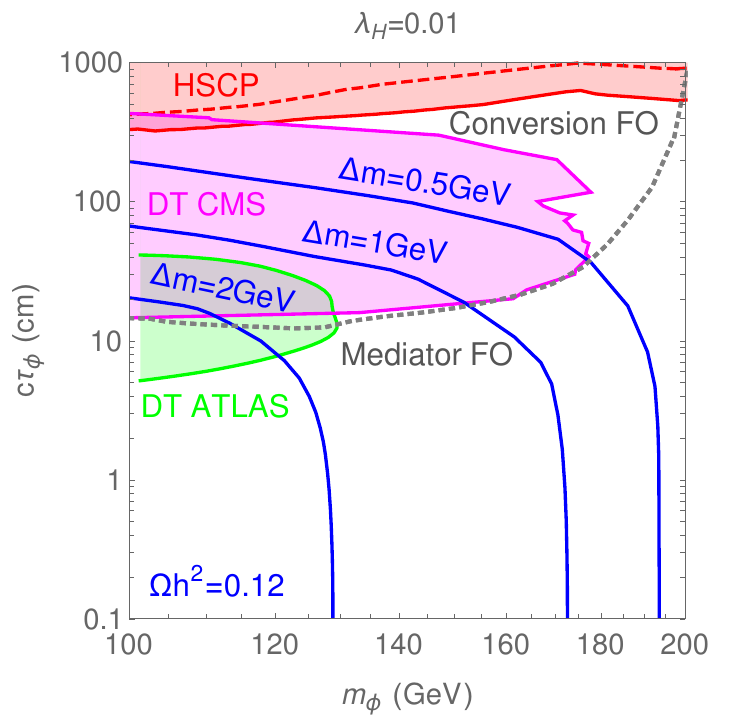}}
	\quad
	\subfloat[$\lambda_H = 0.1$]{\includegraphics[scale=0.6]{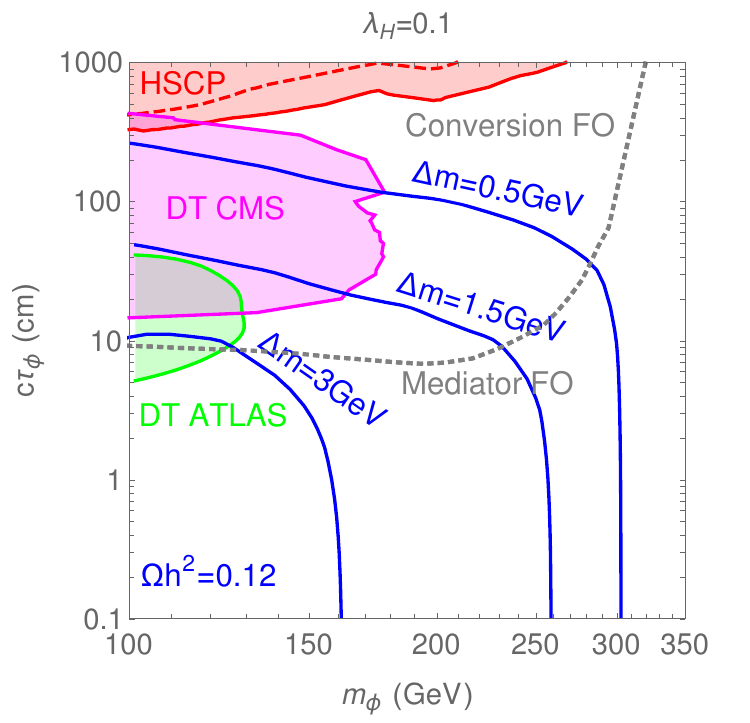}}
	\quad
	\subfloat[$\lambda_H = 0.5$]{\includegraphics[scale=0.6]{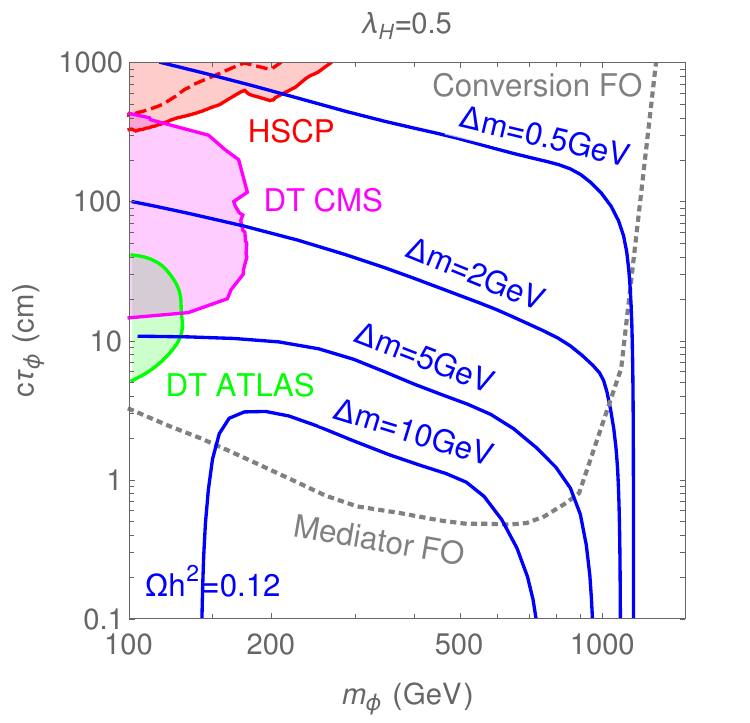}}
	\caption{ Proper life time of the mediator as a function of
          the dark matter mass. Each panel corresponds to a different
          value of $\lambda_H$ contributing to the mediator
          annihilation. The blue contours reproduce the correct relic
          abundance in the $(m_\chi, c\tau_\phi)$ plane for different
          values of the mass-splitting $\Delta m$.  The gray dotted
          line separates the conversion driven (top) from the mediator
          annihilation driven (bottom) freeze-out regime. The excluded
          regions resulting from heavy stable particle (HSCP, red
          region) searches and disappearing tracks (DT, green region)
          searches at LHC are also shown. See text for details.}
          
	\label{fig:ctauLimits}
\end{figure}

In this section we discuss the collider constraints on the conversion
driven regime. The small Yukawa couplings necessary to reproduce the
correct DM relic abundance shown in Fig. \ref{fig:ParamPlot} (blue
contours), together with the mass compression, imply a small decay width
of the charged mediator through the process $\slepton \rightarrow \chi
l$. For $\Delta m\ll m_\chi$, the decay rate for $\slepton \rightarrow
\chi l$ reduces at first orders in $\Delta m$ to:
\begin{equation}
  \Gamma_{\slepton} \approx \frac{\lambda_{\chi}^2 \Delta m^2}{4\pi m_{\chi}} \left[ 1- \frac{2 \Delta m}{m_{\chi}}\right]
  \sim \frac{1}{25 \text{ cm}} \left(\frac{\lambda_{\chi}}{10^{-6}}\right)^2 \left(\frac{\Delta m}{1 \text{ GeV}} \right)^2 \left( \frac{100 \text{ GeV}}{m_{\chi}} \right)
\label{eq:decaysmalldm}
\end{equation}
when neglecting the lepton mass ($m_l \ll m_{\chi},
m_{\slepton}$). The testable signature at colliders for this class of
model is hence the pair production of charged mediators through gauge
interaction and possibly their subsequent macroscopic decay into DM
plus leptons.  
\footnote{The mediator pair production through s-channel Higgs adds up
  to the Drell Yan production considered here. We checked that this
  extra contribution does not qualitatively affect the constraints
  shown in Fig.~\ref{fig:ctauLimits} and hence, we conservatively
  neglect it. } In the next sections we will discuss in details the
possible collider signatures associated to these processes and in
particular the expected sensitivity at the LHC. For a recent overview
of long-lived particle (LLP) searches see~\cite{Alimena:2019zri}.

The overall result of our investigations is shown in
Fig.~\ref{fig:ctauLimits} in the mediator proper decay length $c
\tau_\phi= 1/\Gamma_\phi$ versus mediator mass plane. The blue
contours give rise to $\Omega h^2=0.12$ for fixed values of the mass
splitting.  We consider mediator masses with $m_\phi>100$ GeV as
charged particles with smaller masses are typically excluded by LEP
data~\cite{Abbiendi:2005gc}.  The gray dotted line indicates the
region where we go beyond CE regime.  Above the gray line the DM
abundance can be accounted for through conversion driven freeze-out
while below the gray line, mediator annihilation driven freeze-out is
at work. In the latter case, the relic abundance become essentially
independent of the coupling $\lambda_\chi$ and is determined by a
combination of the parameters $\lambda_H, m_\phi$ and $\Delta m$.  For
fixed values of the latter parameters, several values of
$\lambda_\chi$ (or equivalently $c \tau_\phi$ given
Eq.~(\ref{eq:decaysmalldm})) can thus account for the same DM
abundance. As a result the blue contours become vertical in the bottom
part the plots.  Notice that for e.g. $\lambda_H=0.5$ case the
inverse-$U$-shape of the blue contour at $\Delta m= 10$ GeV is to be
expected given the form of the blue and green contours of the
Fig.~\ref{fig:ParamPlot}. In particular the green contours tell us
that the right abundance can be obtained at $\Delta m= 10$ GeV in the
mediator annihilation freeze-out regime for two dark matter masses:
$m_\chi\approx 150$ and 700 GeV. This is indeed what can be inferred
from the lower region of the $\lambda_H=0.5$ plot of
Fig.~\ref{fig:ctauLimits}.

    Further constraints can be derived exploiting the long lifetime of
    the mediator.  The charged track limit derived from the Heavy
    Stable Charged Particle (HSCP) searches of
    Refs.~\cite{Khachatryan:2015lla} and~\cite{CMS:2016ybj} are
    displayed in red in Fig. \ref{fig:ctauLimits}.  In
    Sec.~\ref{sec:HSCP} we describe the detailed derivation of these
    bounds.  For a moderate decay length of the mediator the relevant
    signature is covered by the disappearing charged tracks (DT)
    searches since the final state particles (lepton plus dark matter)
    are not visible in the detector because of the small mass
    splitting.  The purple region shows the region excluded by the CMS
    analysis \cite{Sirunyan:2018ldc}, while the bound obtained from
    the ATLAS search \cite{Aaboud:2017mpt} is displayed in green.  The
    reinterpretation of these analysis, originally performed for
    long-lived charginos in supersymmetric models, for our DM model is
    discussed in Sec.~\ref{sec:DT}.  Note that, as already noticed in
    e.g.~\cite{Belanger:2018sti}, ATLAS and CMS provide a
    complementary constraints on the lifetime of the mediator.

    We comment on
    displaced leptons searches in Sec.~\ref{sec:displ} and on other
    possibly relevant LHC searches in
    Sec.~\ref{sec:comments-other-lhc}. We will see that with those
    searches we do not get any further constraints on the parameter
    space of Fig.~\ref{fig:ctauLimits}.  It is interesting to observe
    that the current LHC reach on this DM model, even if it includes
    light charged mediators, is very weak, and a large portion of the
    parameter space remains unconstrained.  We comment on possible
    strategies to improve the LHC sensitivity along the discussion of
    the different existing searches.

Finally, let us comment on possible constraints beyond the collider
phenomenology.  First, we notice that the mediator lifetime is always
safely below the BBN constraints, see \cite{Jedamzik:2006xz} for an
estimation.  Furthermore, the leptophilic model considered here with
compressed DM-mediator spectrum could potentially give rise to
enhanced gamma-ray lines and bremsstrhlung relevant for indirect dark
matter searches, direct detection constraints from anapole moment
contributions as well as lepton magnetic moment and Lepton flavour
violation see
e.g.~\cite{Garny:2011ii,Garny:2015wea,Garny:2013ama,Bringmann:2012vr,Giacchino:2013bta,Kopp:2014tsa}. Even
in the WIMP case, only gamma ray constraints and anapole contributions
to the DM scattering on nucleon were shown to marginally constrain the
DM parameter space, see~\cite{Kopp:2014tsa}. In all cases, the
relevant observables scale as $\lambda_\chi$ to the power 2 or
4~\cite{Garny:2011ii,Garny:2015wea,Garny:2013ama,Bringmann:2012vr}. Given
that even smaller dark matter coupling are considered for conversion driven
freeze-out, no extra constraints can be extracted.

\subsubsection{Charged tracks}
\label{sec:HSCP}

One possible avenue to constrain the model under study is to look at
charged tracks resulting from pair production of the mediators that
are sufficiently long lived to decay outside the detector. In order to
reinterpret available analysis within the framework considered here, we
have made use of the public code
{\tt SModelS}~\cite{Kraml:2013mwa,Ambrogi:2018ujg}. This code allows for a
decomposition of the collider signature of a given new physics model
(within which the stability of the DM is ensured by $Z_2$ symmetry)
into a sum of simplified-model topologies. {\tt SModelS} can then use the
cross-section upper limits and efficiency maps provided by the
experimental collaborations for simplified models and apply them to
the new BSM model under study.

We consider the cases in which these topologies are expected to give
rise to 2 HSCP in the final state.\footnote{Notice that currently {\tt
    SModelS} can not constrain displaced vertex (due to mediator
  decays happening within the detector). Such final states are
  currently discarded for the HSCP analysis. The result of the HSCP
  analysis is thus conservative as particles decaying inside the
  detector may provide additional sensitivity.}  For the latter
purposes, {\tt SModelS} evaluates the fraction of BSM particles
decaying outside the detector ${\cal F}_{long}$. The latter are
computed making use of the approximation:
\begin{equation}
  {\cal F}_{long}=\exp\left(-1\frac{1}{c\tau} \left\langle\frac{l_{out}}{\gamma \beta}\right\rangle_{eff}\right)\,,
\label{eq:flong}
\end{equation}
where $\beta$ is the velocity of the LLP, $\gamma = (1 -\beta^2
)^{-1/2}$, $l_{out}$ is the travel length through the CMS detector
(ATLAS analysis are not yet included) and $c\tau$ is the LLP proper
decay length. Here we use $\gamma \beta=2.0$ both for our 13 TeV and 8
TeV reinterpretations, see the Appendix \ref{sec:charged-tracks} for a
discussion motivating such a choice.  In addition, we have to provide
{\tt SModelS} with the production cross-sections of our mediator.  The
production cross section is equivalent to the one of a right handed
slepton pair in a SUSY model.  We took the NLO+NLL cross sections
tabulated by the ``LHC SUSY Cross Section Working Group" which have
been derived using Resummino \cite{Fuks:2013lya}. Notice that the
latter just simply correspond to the LO cross-sections (that can be
obtained with Madgraph) with a $K$-factor correction of roughly $1.5$ at $8$ TeV and
$1.3$ at $13$ TeV.

The regions excluded at 95\% CL obtained, using efficiency maps, are
shown in Fig.~\ref{fig:ctauLimits} with red color. The red dashed
curve delimit the 8 TeV exclusion region using the online material
provided by the CMS collaboration in {\tt
  CMS-EXO-13-006}~\cite{Khachatryan:2015lla} while the 13 TeV
continuous red contour uses {\tt CMS-PAS-EXO-16-036} (with
$12.9~\mathrm{fb}^{-1}$ data)~\cite{CMS:2016ybj}. Notice that our 8
TeV curve gives rise to slightly less constraining limits than the ones
derived in~\cite{Evans:2016zau} for similar scenario ($\tilde \tau$
only) while the 13 TeV ones provide more stringent limit in most of
the mass range  and extend to larger
masses. The 13 TeV data exclude dark matter masses up to $\sim 350$
GeV. Higher mass or equivalently lower production cross-sections can
not currently be constrained.
%

\subsubsection{Disappearing tracks}
\label{sec:DT}

If the decay length of the mediator is still macroscopic but
comparable with the typical size of the detector, another interesting
collider signature can be considered.  In this case, the pair produced
mediators travel a certain portion of the detector and then decay into
dark matter plus leptons.   Hence the signal of such process
is characterized by disappearing tracks. Indeed, the dark matter is invisible and,  because of the typical small mass
splitting, the emitted leptons are too soft to be reconstructed.\footnote{In the next subsection we will
  discuss the case of larger mass splitting and investigate whether
  other searches can be effective.}  Similar searches have been
performed at 13 TeV by CMS \cite{Sirunyan:2018ldc} and ATLAS
\cite{Aaboud:2017mpt,ATL-PHYS-PUB-2017-019}, focusing on the case of Wino or Higgsino DM in
supersymmetry.  
  In such cases, the lightest chargino and
neutralinos are almost degenerate in mass.  When the chargino is
produced (in pair or in association with another neutralino) it flies
for a macroscopic distance in the detector and afterwards decays into
the lightest neutralino plus a soft pion. These searches hence target
charged tracks leaving some hits in the detector but not reaching the
outer part, i.e. disappearing tracks.

Our model, for moderate lifetime of the mediator, gives rise to such
signatures at colliders. An important difference with respect to the
SUSY case, usually considered in experimental searches, is that in our
model the number of produced disappearing tracks per event is always
two. This differs with respect to the SUSY case where there can be
chargino pair production but also chargino-neutralino associated
production.

Here we assess the reach of disappearing track searches
on our DM model for both the ATLAS \cite{Aaboud:2017mpt} and the CMS
\cite{Sirunyan:2018ldc} search.  In the auxiliary HEPData material of
the ATLAS search, efficiency maps are provided as a function of the
mass of the long-lived charged particle and its lifetime, for the case
of electroweak production. In Appendix \ref{sec:CollApp}, we first
make use of these efficiencies to reproduce the exclusion curve of the
ATLAS paper for the pure Wino, and we then explain how we derive the
corresponding exclusion curve for our DM model.  In the auxiliary
material of the CMS search, on the other hand, the upper limit cross
section as a function of the Wino mass and of its lifetime is
reported.  In Appendix \ref{sec:CollApp} we explain how (and under
which approximations) we employ this information to derive the
corresponding upper limit on our model.  Notice that we assume that
the efficiencies considered in the ATLAS and CMS analysis respectively
also apply to our model. This is because our mediators are produced
through electroweak processes, like the Wino. In
Fig.~\ref{fig:ctauLimits} we show with a green (purple) region the
exclusion limit that we obtain by reinterpreting the ATLAS (CMS)
disappearing track search on our model assuming a mass splitting such
that the emitted lepton cannot be reconstructed.

\subsubsection{Displaced lepton pairs}

\label{sec:displ}

\begin{figure}
	\centering
	\subfloat[]{\label{fig:pTDistr_8TeV}{\includegraphics[scale=0.55]{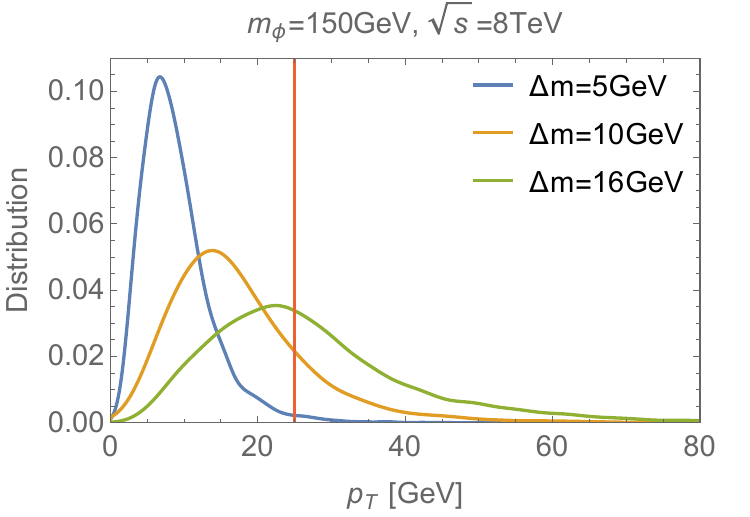}}}
	~~
	\subfloat[]{\label{fig:pTDistr_13TeV}{\includegraphics[scale=0.55]{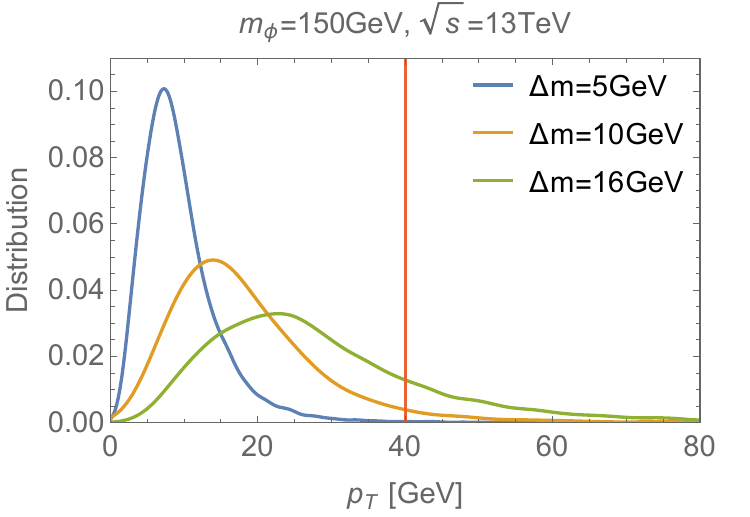}}}
	\caption{The $p_{T}$-distribution of the leading muons for the case when $m_{\phi}=$ 150 GeV and different values of the mass splitting. The red vertical line denotes the cut on the $p_T$ that is made in the search \cite{Khachatryan:2014mea,CMS:2016isf} at 25GeV for $\sqrt{s}=$ 8 TeV (left) and 40 GeV for $\sqrt{s}=$ 13 TeV (right).}
	\label{fig:pTDistr}
\end{figure}
There is actually a significant portion of the parameter space where
the decay of the charged mediators can occur inside the collider, see
Fig. \ref{fig:ctauLimits}.  In the latter case, the final topology of
the signal includes a pair of displaced leptons and a pair of dark
matter particles. If the leptons can be reconstructed, this final
state could be constrained by displaced lepton searches
\cite{Khachatryan:2014mea,CMS:2016isf}. It is important to note that
the CMS search \cite{Khachatryan:2014mea,CMS:2016isf} targets $e \mu$
final states, while in order to probe our model a search targeting
same flavour leptons, as also suggested in \cite{Evans:2016zau}, would
be needed.

Assuming that a same flavour lepton search could be performed, the
focus of our study is anyway in the compressed spectrum regime, and
hence the leptons will be generically soft and difficult to
reconstruct.  Note that the missing energy of the process projected in
the transverse plane will also be negligible because the compressed
mass spectrum implies that the two dark matter particles are mostly
produced back to back.

Nevertheless, in order to estimate the possible reach of a displaced
lepton search on the parameter space of conversion driven freeze out,
we analyze the $p_T$ distribution of the produced lepton on few
benchmarks.  We employ {\tt
  Madgraph}~\cite{Alwall:2014hca,Alwall:2011uj} to simulate the
production of a pair of mediators and their subsequent decay into muon
and dark matter.  In Fig. \ref{fig:pTDistr} we display the $p_T$
distribution for the leading muon for three different mass splittings
$\Delta m = 5,10,16$ GeV.  We note that in
\cite{Khachatryan:2014mea,CMS:2016isf} a minimal $p_T$ cut on the
outgoing muon is 25 and 40 GeV, respectively at $\sqrt{s}=$ 8 TeV and
$\sqrt{s}=$ 13 TeV.  We take these as reference value cuts for an
hypothetical experimental search looking for same flavour displaced
leptons, and we highlight these cuts in Fig.~\ref{fig:pTDistr} with red
vertical lines. From these distributions we conclude that, due to the
small mass splitting, most of the events will not pass the minimal cut
and therefore these searches will not put any constraints on our
model.\footnote{Note also that sizeable values for $\Delta m$
  correspond to large mediator mass (see Fig. \ref{fig:ParamPlot}) and
  hence small production cross sections.}  In this perspective, it
would be interesting to explore how much the minimal $p_T$ cut on the
leptons could be reduced in displaced lepton searches, as also suggested in~\cite{Filimonova:2018qdc}.

\subsubsection{Comments on other LHC searches}
\label{sec:comments-other-lhc}

The difficulties in probing soft objects at the LHC is not only
related to their reconstruction, but also to the necessity of
triggering on the event.  These issues can be typically solved by
considering processes accompanied by energetic initial state QCD
radiation.  In this perspective, relevant searches for our model can
be the monojet analysis \cite{Aaboud:2017phn,Sirunyan:2017hci} or the
recent boosted soft lepton analysis \cite{Aaboud:2017leg}.  However,
the latter does not apply to our scenario since it requires soft
leptons that are promptly produced.

We can easily provide an estimation of the monojet search bound and
conclude that there is not enough sensitivity so as to be relevant for
our model. This is principally due to the small electroweak production
cross section.  As an illustrative case, we considered a benchmark
with $m_{\phi}=100$ GeV for which we simulated the pair production of
mediators with the addition of one extra energetic jet using {\tt
  Madgraph}. The signal region categories of \cite{Aaboud:2017phn}
starts at $E^{miss}_{T} >$ 250 GeV and put 95\% CL on the visible
cross section in several bins with increasing $E^{miss}_{T}$ cuts.  We
have verified that, once an extra jet with the corresponding $p_T$ is
required, the cross section in our model is always at least two order of
magnitudes below the experimentally excluded cross section in all the
$E^{miss}_{T}$ bins.  Hence we conclude that monojet searches can not
constraint the parameter space of our model.

\section{ Conclusions }
\label{sec:concl}

The main goal of the paper is to determine the viable parameter space
and the associated experimental constraints for a dark matter
candidate that account for all the dark matter when produced at early
times through conversion driven freeze-out, i.e. investigating the
dark matter freeze-out beyond CE. We have worked in a simplified
model including a Majorana dark matter fermion $\chi$ coupling to SM
lepton and a charged dark scalar $\phi$, the mediator, through a
Yukawa coupling. The minimal set of extra parameters for this model
are the Yukawa coupling $\lambda_\chi$, responsible for the conversion
processes, the dark matter mass $m_\chi$ and the mass splitting
between the mediator and the DM $\Delta m$. In addition, given that
the mediator is a charged scalar, one can always write a quartic
interactions between $\phi$ and the Higgs driven by the coupling
$\lambda_H$. We have considered a few benchmark values of the latter
coupling for illustration. The parameter space studied here involves
relatively feeble conversion couplings $\lambda_\chi$ compared to
previous analysis made within the context of the WIMP
paradigm~\cite{Garny:2011ii,Garny:2015wea,Garny:2013ama,Bringmann:2012vr,Giacchino:2013bta,Kopp:2014tsa}
and extend the conversion driven
analysis~\cite{Garny:2017rxs,Garny:2018icg} to the case of a
leptophilic scenario.

As a first step, we have described how varying the conversion
coupling one can continuously go from DM annihilation driven
freeze-out to freeze-in passing through mediator annihilation and
conversion driven freeze-out in the context of compressed mass
spectrum. Our discussion is summarized in Fig.~\ref{fig:ann-conv}.  We
have then focused on the case of conversion driven freeze-out.
We have first
determined the viable parameter space in order to account for all the
DM that typically involves conversion couplings in the range
$\lambda_\chi\in [10^{-7},10^{-6}]$. For negligible $H$-$\phi$
coupling $\lambda_H= 0.01$, the latter reduces to a limited parameter
space with $m_\chi<200$~GeV and $\Delta m <2.6$~GeV. Larger
$\lambda_H$ increase the mediator annihilation cross-section and, by
the same token, extend the viable parameter space to e.g.
$m_\chi<1$~TeV and $\Delta m <14$~GeV for $\lambda_H=0.5$~GeV. The
precise viable parameter space is shown Fig.~\ref{fig:ParamPlot} for
a selection of $\lambda_H$ coupling. 

Finally we have addressed the collider constraints on the model under
study. Interestingly the feeble coupling involved give rise to a long
lived mediator that can a priori be tested through existing searches
for heavy stable charged particles, disappearing charged tracks and,
possibly, displaced leptons. We have recasted existing searches and
projected the results within our scenario as shown in
Fig.~\ref{fig:ctauLimits}. As can be seen, only disappearing charged
tracks and heavy stable charged particle searches provide relevant
constraints on the parameter space. We also explicitly checked that
displaced leptons and monojet searches can not help to further test
our DM scenario. At this point, a large part of the parameter space is
left unconstrained.

\section*{Acknowledgments}
We thank Lorenzo Calibbi, Francesco D'eramo, Nishita Desai, Marie-H\'el\`ene Genest, Jan Heisig, Andre
Lessa, Zachary Marshall, Diego Redigolo, Ryu Sawada, J\'er\^ome Vandecasteele, Laurent
Vanderheyden, Matthias Vereecken and Bryan Zaldivar for useful
discussions.  SJ is supported by Université Libre de Bruxelles PhD
grant and LLH is a Research associate of the Fonds de la Recherche
Scientifique FRS-FNRS. This work is supported by the FNRS research
grant number F.4520.19; by the Strategic Research Program
\textit{High-Energy Physics} and the Research Council of the Vrije
Universiteit Brussel; and by the ``Excellence of Science - EOS'' -
be.h project n.30820817.

 \appendix

\section{Relevant processes for the Relic abundance computation}
\label{sec:processes}

In table \ref{tab:coann} and \ref{tab:conv}, all the processes that
influence the relic abundance are summarized. All of them have a
different influence and therefore, we compare all the relevant rates of
interaction to the Hubble rate. If $\Gamma>H$, they are efficient and
the interactions happen fast enough such that the process is in
equilibrium.  The rates used in Fig.~\ref{fig:Rateplot3} and \ref{fig:Rateplot1} are defined as $\Gamma_{ij\to k(l)}=
  \gamma_{ij\to k(l)}/n^{eq}_{\chi}$, except for the rate of $\phi$
  annihilation in which case $\Gamma_{\phi\phi^\dag\to {\rm SM\,SM}}=
  \gamma_{\phi\phi^\dag}/n^{eq}_{\phi}$. If one compares the rates in Fig.~\ref{fig:Rateplot3} and \ref{fig:Rateplot1} with tables \ref{tab:coann} and
\ref{tab:conv}, one can spot that there is one process missing, namely
$\slepton \slepton \rightarrow ll$. This process has the same
influence as $\slepton \slepton^{\dagger} \rightarrow SM SM$ on the
relic abundance, but it is suppressed by $\lambda_{\chi}^4$. It is
always sub-leading (unless $\lambda_{\chi} \approx 1$, a regime we are
not interested in) and therefore, it is not included in the plots where
we compare the efficiencies of the different rates or in the Boltzmann
equation. \\
 The decay rate for $\slepton
\rightarrow \chi l$ (not thermally averaged!) reads 
\begin{align}
\label{eq:decayRate}
	\Gamma_{\slepton} &= \lambda_{\chi}^2 \frac{(m_{\slepton}^2-m_l^2-m_{\chi}^2) \sqrt{\left[m_{\slepton}^2-(m_{\chi}-m_l)^2\right] \left[m_{\slepton}^2-(m_{\chi}+m_l)^2\right]}}{16 \pi m_{\slepton}^3},\\
	&\approx \frac{\lambda_{\chi}^2 \Delta m^2}{4\pi m_{\chi}} \left[ 1- \frac{2 \Delta m}{m_{\chi}}+\dots \right],
\end{align}
where in the second line we have assumed small mass splittings
($\Delta m\ll m$) and we have neglected the lepton mass.  The
cross-sections have been obtained making use of {\tt FeynRules}
\cite{Alloul:2013bka} and {\tt Calchep}\cite{Belyaev:2012qa} to
extract the transition amplitudes $ \mathcal{M}$ as:
\begin{align}
	\langle\sigma_{ij}v\rangle n_{i,eq} n_{j,eq} = \frac{g_i g_j}{512 \pi^5} T \int \frac{|\mathcal{M}|^2}{\sqrt{s}} K_1 \left( \frac{\sqrt{s}}{T} \right) ds \ dt,
\end{align}
with
\begin{equation}
n_{i,eq}= \frac{g_i}{2 \pi^2} 
m_i^2 T K_2 (m_i /T ),
\end{equation}
where $ K_{1,2}$ are the first and second modified Bessel functions of the 2nd
kind. In some of the above cases,
some $s-$ and $t-$channel divergences can appear. In the latter cases,
we follow the same procedure as in~\cite{Garny:2017rxs} introducing
cuts in the integration regions to handle them. We checked that our
results were stable varying the cuts.

\begin{table}
	\centering
	\renewcommand\arraystretch{1.5}
	\begin{tabular}{| c | c | c | c | c | }
		\hline 
		\multicolumn{2}{|c|}{initial state} & \multicolumn{2}{|c|}{final state} & scaling with  $\lambda_{\chi}$ \\
		\hline	\hline
		$\chi$ & $\chi$ & $l^-$ & $l^+$ & $\lambda_{\chi}^4$ \\
		\hline	\hline
		\multirow{2}{*}{$\chi$} &  \multirow{2}{*}{$\slepton$}
		& $l^-$ & $\gamma,Z,H$ & \multirow{2}{*}{$\lambda_{\chi}^2$} \\
		\cline{3-4}
		& & $W^-$ & $\nu_{l}$ & \\
		\hline \hline
		\multirow{4}{*}{$\slepton$} &  \multirow{4}{*}{$\slepton^{\dagger}$}
		& $\gamma,Z,W^+$ & $\gamma,Z,W^-$ & \multirow{4}{*}{$\lambda_{\chi}^0$} \\
		\cline{3-4}
		& & $q$ & $\bar{q}$ & \\
		\cline{3-4}
		& & $H$ & $Z,H$ & \\
		\cline{3-4}
		& & $l^-$ & $l^+$ & \\
		\hline
		$\slepton$ & $\slepton$ & $l^-$ & $l^-$ & $\lambda_{\chi}^4$\\
		\hline
	\end{tabular}
	\caption{List of all relevant (co-)annihilation processes. The
            leptons involved are denoted by $l =e,\mu,\tau$. Also the
            dependence of the cross section on the coupling constant
            $\lambda_{\chi}$ is denoted in the last column. The
            $\slepton \slepton^{\dagger}$ annihilation into $l\bar{l}$
            also has contributions scaling with $\lambda_{\chi}^2$ and
            $\lambda_{\chi}^4$.  }
	\label{tab:coann}
\end{table}
\begin{table}
	\centering
	\renewcommand\arraystretch{1.5} 
	\begin{tabular}{| c | c | c | c | c | }
		\hline 
		\multicolumn{2}{|c|}{initial state} & \multicolumn{2}{|c|}{final state} & scaling \\
		\hline	\hline
		\multirow{4}{*}{$\chi$} & $l^-$ & \multirow{4}{*}{$\slepton$} & $\gamma,Z,H$ & \multirow{4}{*}{$\lambda_{\chi}^2$} \\
		\cline{2-2} \cline{4-4}
		& $\gamma,Z,H$ & & $l^+$ & \\
		\cline{2-2} \cline{4-4}
		& $W^-$ & & $\bar{\nu_l}$ & \\
		\cline{2-2} \cline{4-4}
		& $\nu_l$ & & $W^+$ & \\
		\hline \hline
		\multicolumn{2}{|c|}{$\slepton$} & $\chi$ & $l^-$ & $\lambda_{\chi}^2$\\
		\hline \hline
		$\chi$ & $\chi$ & $\slepton$ & $\slepton^{\dagger}$ & $\lambda_{\chi}^4$ \\
		\hline
	\end{tabular}
	\caption{List of all included conversion processes and their dependence of the cross section on $\lambda_{\chi}$. $l$ is one of the leptons ($e,\mu,\tau$), depending on the case we are studying.}
	\label{tab:conv}
\end{table}

\begin{figure}
	\centering
	\includegraphics[scale=0.75]{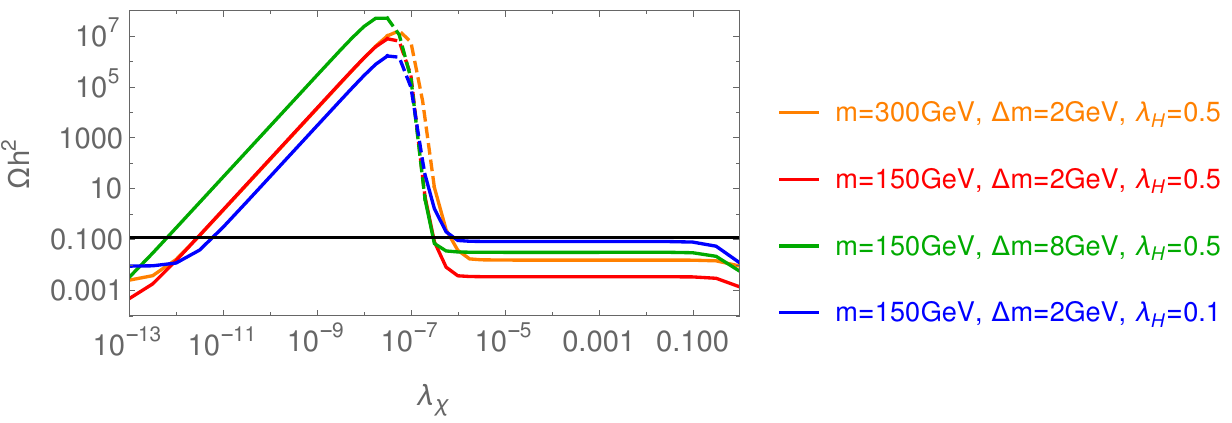}
	\caption{DM abundance as a function of the Yukawa coupling (as in Fig. \ref{fig:ann-conv}) for different values of the parameters $m_\chi$, $\Delta m$ and $\lambda_H$}
	\label{fig:multiLambdaPlot}
\end{figure}

Finally, when discussing the dependence of the relic abundance on the
conversion coupling we looked at a particular benchmark with a
coupling to the SM muon and $m_{\chi} = 150$ GeV, $\Delta m = 2$ GeV,
and $\lambda_{H} = 0.1$. The latter case was shown in
Fig.~\ref{fig:ann-conv}. For the sake of illustration we also show how
the $\Omega h^2$ curve is affected varying $m_{\chi}$, $\Delta m$, and
$\lambda_{H}$ in Fig.~\ref{fig:multiLambdaPlot}, see also the
discussion in Sec.~\ref{sec:viab}.

\section{Dependence on initial conditions }
\label{sec:InitCond}
\begin{figure}
	\centering
	\subfloat[$\lambda_{\chi} = 4 \cdot 10^{-7}$]{\label{fig:InitCondFullA}{\includegraphics[scale=0.55]{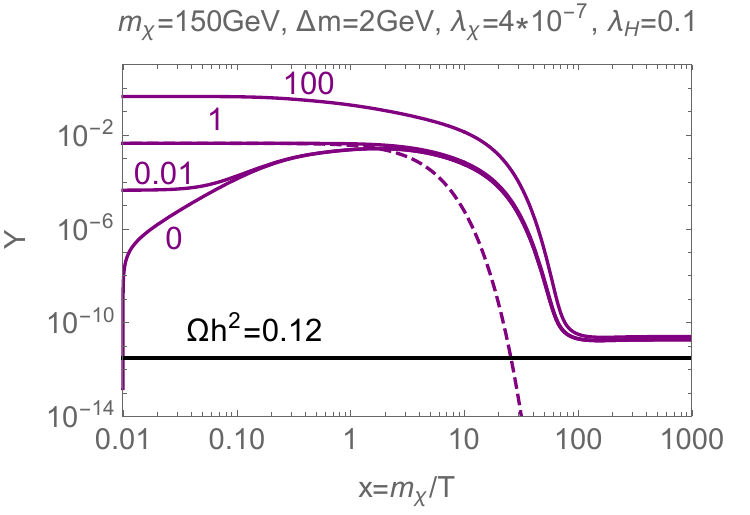}}}
	~~
	\subfloat[$\lambda_{\chi} = 10^{-7}$]{\label{fig:InitCondFullB}{\includegraphics[scale=0.55]{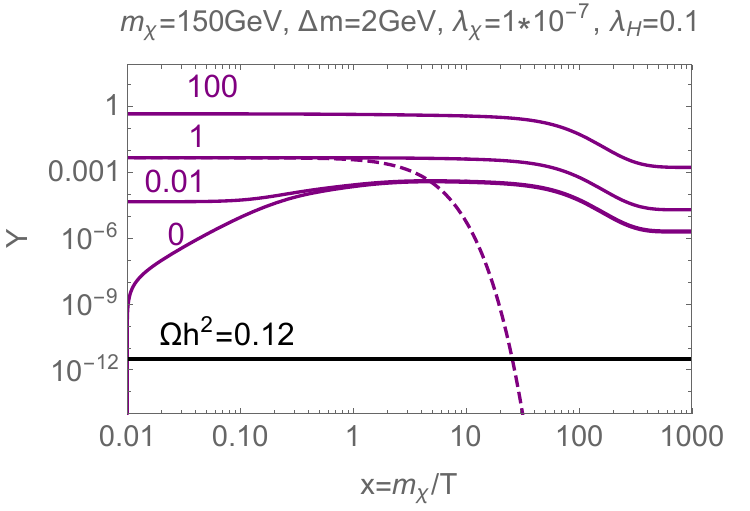}}}
	\caption{Plotting the evolution of the yield for different initial conditions: $Y_{\chi}[1]= i \cdot Y_{\chi,eq}$[1] with $i \in \{ 0,0.01,1,100 \}$. Both figures are calculated with $\smuon$ as the co-annihilation partner and for the parameters $m_{\chi} = 150$ GeV and $\Delta m = 2$ GeV but for two different values of the coupling, $\lambda_{\chi} = 4 \cdot 10^{-7}$ (left) and $\lambda_{\chi} = 10^{-7}$ (right). The dashed lines denote the equilibrium yield.}
	\label{fig:InitCondFull}
\end{figure}

A very interesting feature of the freeze-out mechanism is that it does
not depend on the initial conditions. For small values of $x$ ($x
\lesssim 1$), the annihilation rates are efficient and therefore, the
yield always converges to the equilibrium value, regardless whether we
start with zero or with a very high abundance. Instead, in the case of small
coupling involved, as the freeze-in case, it is well known that the
final DM abundance can be sensitive to the initial conditions
(IC). In~\cite{Garny:2017rxs}, it was pointed out that when DM couples
to light quarks through Eq.~(\ref{eq:lagr}), the conversion driven
abundance is independent of the IC. In contrast, in the case of a
coupling to leptons considered here, we clearly notice a dependence on
the IC, especially for values of $\lambda_{\chi} \sim \mathcal{O}(10^{-7})$ or less. 
This is illustrated in Fig.~\ref{fig:InitCondFull} where we show
the yields for 2 distinct values of the Yukawa coupling. The reason
for this dependence in the leptophilic case is due to the fact that
conversion driven processes are less efficient than in the
quark-philic case as the gauge coupling  involved in the processes depicted
in Fig.~\ref{fig:feyn_conv}  is the EW coupling  instead of strong coupling. In the
follow-up we will always assume that $Y_{\chi}(0.01)=0$.

\section{Technical details on the collider searches}
\label{sec:CollApp}

\subsection{Charged Tracks}
\label{sec:charged-tracks}

A priori, the evaluation of ${\cal F}_{long}$ in Eq.~(\ref{eq:flong})
would require an event-based computation of $l_{out}/\gamma
\beta$. In the appendix of \cite{Heisig:2018kfq}, it is however argued
that considering the effective length $L_{eff}=
\left\langle{l_{out}}/{\gamma \beta}\right\rangle_{eff}=7 m$ provide
conservative constraints. This $L_{eff}$ corresponds to a typical
$\gamma \beta \simeq 1.43$ for $l_{out}\simeq 10$ m and is encoded by
default in the SModelS code.\footnote{ The $L_{eff}=7$ m or more
  precisely $\gamma \beta \simeq 1.43$ is encoded into the file {\tt
    smodels/theory/slhaDecomposer.py}} A closer look to their figure
B.6. (in the case of the LLP direct production that we consider here)
already tells you that in the mass range of a few hundred of GeV LLP
mass, this $\gamma \beta \simeq 1.43$ give rise to a weaker constraint
on the LLP life time than in the event-based computation. Making use
of our own Madgraph simulations we obtain that the resulting $\gamma
\beta$ distribution tends to  values larger than 1.43. In particular for a
mediator mass between 100 and 250 (100 and 350) GeV, we obtain a mean
$\gamma \beta$ varying between 3.5 and 2.0 (4.9 and 2.17) for
$\sqrt{s}= 8$ TeV (13 TeV). As a result, here we use the conservative
value of $\gamma \beta=2.0$ (or equivalently $L_{eff}=5$ m) for our
analysis.

\subsection{Disappearing Tracks}
\subsubsection*{ATLAS Search}
\begin{figure}
  \centering
  \includegraphics[scale=0.8]{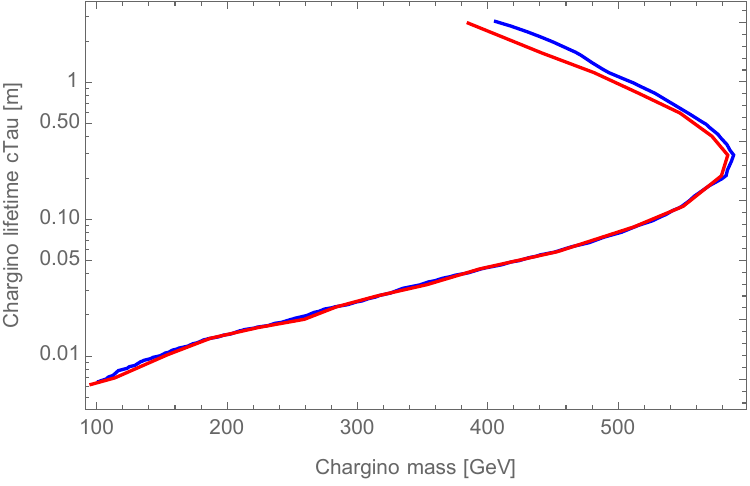}
  \caption{Exclusion limit at 95\% CL for the supersymetric pure-wino chargino/neutralino analysis obtained from ref. \cite{Aaboud:2017mpt} (red) and by using the recasting of the efficiencies as explained in the text (blue).}
  \label{fig:WinoExcl}
\end{figure}

The ATLAS search \cite{Aaboud:2017mpt} for events with at least one disappearing track focussed on the case of wino DM. In these supersymmetric models, the lightest chargino and neutralino are almost pure wino and they are nearly mass degenerate. Therefore, the chargino can decay to a neutralino and a soft pion inside the detector, leaving a disappearing track. The ATLAS collaboration provided the efficiency maps for this search in the HEPData. In order to use these efficiency maps to asses the reach of disappearing track searches on our DM model, we need to know how to interpret them.\\ 
Besides the event acceptance $E_A$ and efficiency $E_E$, the tracklet\footnote{A tracklet is a track in the detector between 12 and 30 cm.} too has to pass the reconstruction selection requirements. The probability to pass the generator-level requirements is the tracklet acceptance $T_A$. The probability to pass the full pixel tracklet selection at reconstruction level is the tracklet efficiency $T_E$. Both should be applied for every tracklet. Finally, the probability for a tracklet to have a $p_T > 100GeV$ is denoted independently by $P$ and is taken to have a constant value of 0.57 for the charginos. \\
There are three different processes that can leave a disappearing track in the detector for the wino-like chargino/neutralino model,
\begin{align}
 & p p \rightarrow \chi^+_1 \chi^-_1,  \label{poc:2tracks}\\
 & p p \rightarrow \chi^+_1 \chi^0_1, \label{poc:1tracks1}\\
 & p p \rightarrow \chi^-_1 \chi^0_1, \label{poc:1tracks2}
\end{align}
with each of the processes having approximately the same production cross section, i.e. one third of the total cross section. The efficiency map provided in the HEPData\footnote{We thank the ATLAS exotics conveners for information about the efficiency maps in the HEPData.} denotes the total model dependent efficiency (i.e. taking into account the fact that some of the above processes can leave two tracks in the detector), without taking into account the probability $P$ (it will be reintroduced later). Since in our model, we have only processes that can leave two tracks, we need to obtain the efficiency for two tracklet processes. \\
In general, the probability $\mathcal{E}_N$ of reconstructing at least one tracklet coming from a process leaving $N$ tracks in the detector has an efficiency of,
\begin{align}
   \mathcal{E}_N &= E_A \times E_E \times (1-(1-T_A \times T_E )^N) \nonumber \\
   &\approx E_A \times E_E \times (1-(1-N \, T_A \times T_E)) \nonumber \\
  &= N E_A \times E_E \times T_A \times T_E \nonumber \\
  &= N \mathcal{E}_1.
  \label{eq:effNtracks}
\end{align}
For the wino-like chargino/neutralino analysis, only process (\ref{poc:2tracks}) can leave two tracks in the detector while the other two can only leave one. Therefore, the model dependent efficiency for the pure wino chargino/neutralino analysis is
\begin{align}
  \mathcal{E}_{full} = \frac{2}{3} \mathcal{E}_1 + \frac{1}{3} \mathcal{E}_2 \approx \frac{4}{3} \mathcal{E}_1 \approx  \frac{2}{3} \mathcal{E}_2,
  \label{eq:FullEff}
\end{align}
where the $2/3$ represents the contribution from the processes with one tracklet (i.e. \eqref{poc:1tracks1} and \eqref{poc:1tracks2}) and the $1/3$ represents the contribution from the processes with two tracklets \eqref{poc:2tracks},
all assumed to have the same production cross section.
As mentioned, the full efficiency $ \mathcal{E}_{full}$ is the one reported in the HEPData and we use Eq. (\ref{eq:effNtracks}) and (\ref{eq:FullEff}) to derive the efficiency for a two tracklet process, that we use in the analysis of our DM model. \\
In order to validate this technique, we reproduce the exclusion limit at 95\% CL for the pure wino case \cite{Aaboud:2017mpt} in figure \ref{fig:WinoExcl}, by multiplying the full efficiency with the total cross-section, together with the probability $P$, to get the visible cross-section.

\subsubsection*{CMS Search}
The CMS search for disappearing tracks \cite{Sirunyan:2018ldc} focuses on the same model as the ATLAS search, i.e. wino DM. In the HEPData, they provide a 95\% CL upper limit $\sigma_{Wino}^{UL} $ on the product of the cross section for direct production of charginos as a function of chargino mass and lifetime. Since this is a model dependent limit, we have to recast this upper limit to account for the fact that for the model under study here, we will always produce two charged tracks in the detector. In order to accommodate this in our analysis, we will make use of the same approximations as for the ATLAS search, namely that the efficiency used for the search can be approximated by Eq. (\ref{eq:FullEff}). Together with this approximation, we can obtain a 95\% CL exclusion limit on the charged scalar $\phi$ in the model under study here from the data given on the HEPData page of the experiment as follows
\begin{equation}
	\sigma_{\phi}^{UL} = \frac{\sigma_{vis}^{UL}}{\mathcal{E}_2} \approx \frac{2}{3} \frac{\sigma_{vis}^{UL}}{\mathcal{E}_{full}} = \frac{2}{3} \sigma_{Wino}^{UL}.
\end{equation}

%
\bibliographystyle{JHEP} 
\bibliography{bibSD-FI}
%
%
\end{document}